\documentclass[pra,twocolumn,preprintnumbers,amsmath,amssymb,nofootinbib,floatfix]{revtex4}

\usepackage{graphicx,bm, tikz}
\usepackage[breaklinks]{hyperref}

\makeatletter
\def\graphicscale{\twocolumn@sw{0.3}{0.4}}
\def\graphicthreescale{\twocolumn@sw{0.3}{0.4}}

\begin{document}

\title{Dynamics after quenches in one-dimensional quantum Ising-like systems}

\author{Davide Rossini}
\affiliation{Dipartimento di Fisica dell'Universit\`a di Pisa
        and INFN, Largo Pontecorvo 3, I-56127 Pisa, Italy}

\author{Ettore Vicari} 
\affiliation{Dipartimento di Fisica dell'Universit\`a di Pisa
        and INFN, Largo Pontecorvo 3, I-56127 Pisa, Italy}

\date{\today}

\begin{abstract}
  We study the out-of-equilibrium dynamics of one-dimensional quantum
  Ising-like systems, arising from sudden quenches of the Hamiltonian
  parameter $g$ driving quantum transitions between disordered and
  ordered phases.  In particular, we consider quenches to values of
  $g$ around the critical value $g_c$, and mainly address the question
  whether, and how, the quantum transition leaves traces in the
  evolution of the transverse and longitudinal magnetizations during
  such a deep out-of-equilibrium dynamics.  We shed light on the
  emergence of singularities in the thermodynamic infinite-size limit,
  likely related to the integrability of the model.  Finite systems in
  periodic and open boundary conditions develop peculiar power-law
  finite-size scaling laws related to revival phenomena, but
  apparently unrelated to the quantum transition, because their main
  features are generally observed in quenches to generic values of
  $g$. We also investigate the effects of dissipative interactions
  with an environment, modeled by a Lindblad equation with local decay
  and pumping dissipation operators within the quadratic fermionic
  model obtainable by a Jordan-Wigner mapping.  Dissipation tends to
  suppress the main features of the unitary dynamics of closed
  systems. We finally address the effects of integrability breaking,
  due to further lattice interactions, such as in anisotropic
  next-to-nearest neighbor Ising (ANNNI) models. We show that some
  qualitative features of the post-quench dynamics persist, in
  particular the different behaviors when quenching to quantum
  ferromagnetic and paramagnetic phases, and the revival phenomena due
  to the finite size of the system.
\end{abstract}

\maketitle

% ========================= BODY =========================

\section{Introduction}
\label{intro}

The quantum evolution of many-body systems has been considered a
challenging problem for long time.  The recent experimental progress in the
realization, control, and readout of the coherent dynamics of isolated
quantum many-body systems (see e.g. Refs.~\cite{Bloch-08, GAN-14}) has
made this issue particularly relevant for experiments and
realizations of physical devices for quantum computing.

The simplest protocols in which the out-of-equilibrium dynamics of
many-body systems can be investigated are provided by the so-called
{\it quantum quenches}~\cite{Greiner-02, Weiss-06, Schmiedmayer-07,
  Trotzky-12, Cheneau-12, Schmiedmayer-12}.  A quench protocol is
generally performed within Hamiltonians that may be written as the sum
of two noncommuting terms: $\hat H(g) = \hat H_u + g \hat H_g$, with a
tunable parameter $g$. In practice, one may start from the ground
state $|\Phi_{g_0}\rangle$ of the Hamiltonian $\hat H(g_0)$ associated
with an initial value of the parameter $g_0$,
i.e.~$|\Psi(t=0)\rangle = |\Phi_{g_0}\rangle$, and then suddenly change
it to $g \neq g_0$.  Then the quantum evolution gets driven by the
Hamiltonian $\hat H(g)$, that is,
\begin{equation}
  |\Psi(t)\rangle = e^{-i \hat H(g) t} |\Phi_{g_0}\rangle
  \label{unitary}
\end{equation}
(hereafter we adopt units of $\hbar = 1$).  Several interesting issues
have been investigated for the quantum dynamics after a quench.  They
include the long-time relaxation and the consequent spreading of
quantum correlations and entanglement, the statistics of the work,
localization effects due to the mutual interplay of interactions and
disorder, dynamical phase transitions, the dynamic scaling close to
quantum transitions, effects of dissipation due to interactions with
an environment, the relevance of quantum measurements protocols after
quenches at quantum transitions, to mention some of the most
representative ones (see, e.g., Refs.~\cite{Niemeijer-67,BMD-70,
  DMCF-06, RDYO-07, RDYO-08, ZPP-08, PZ-09, PSSV-11, IR-11, RI-11, GS-12,
  CEF-12-1,CEF-12-2, BRI-12, CE-13, FCEC-14, NH-15, CTGM-15,
  CC-16, BD-16, IMPZ-16, LGS-16, VM-16, NRVH-17, Heyl-17, PRV-18b,
  NRV-19-w, NRV-19, STT-19, RV-20-qm} and references therein).

Quantum phase transitions are striking signatures of many-body
collective behaviors~\cite{SGCS-97, Sachdev-book}.  They are
essentially related to the equilibrium low-energy properties of the
system. However, they could also be probed by out-of-equilibrium
dynamic protocols, for example analyzing the effects of slow changes
of the Hamiltonian parameters across them~\cite{Kibble-76, Zurek-85,
  ZDZ-05,PG-08, CEGS-12, RV-20}.  Recently, some studies have also
focused on the possibility of probing quantum transitions analyzing
the out-of-equilibrium dynamics arising from quantum quenches, for
example when instantaneously changing the Hamiltonian parameters,
setting them to values corresponding to different quantum phases (see,
in particular, Refs.~\cite{BDD-15, RMD-17, TIGGG-19, HPD-18,
  HMHPRD-20}).

The out-of-equilibrium dynamics at the quantum transitions develops
scaling behaviors controlled by the universality class of the quantum
transition, located at a given $g = g_c$. Indeed such dynamic scaling
is observed when the Hamiltonian parameters are slowly changed across
the transition~\cite{PSSV-11, CEGS-12, RV-20}, and also at quantum
quenches when both Hamiltonian parameters associated with the quench,
$g_0$ and $g$, are close to the quantum critical point~\cite{PRV-18b,
  PRV-20}.  However, in the case of hard quenches, i.e. when $g_0$ and
$g$ differ significantly, a dynamic scaling controlled by the
universality class of the equilibrium quantum transition is not
expected, essentially because the instantaneous change of the
Hamiltonian parameters entails a significant amount of energy
exchange. Indeed, the instantaneous change from $g_0$ to $g$ gives
rise to a relatively large amount $\Delta E$ of energy above the
ground level of the Hamiltonian $\hat H(g)$,
\begin{equation}
  \Delta E = \sum_{n} [E_{n}(g) - E_{0}(g)] 
|\langle n,g| \Phi_{g_0} \rangle|^2 \sim L^d\,,
  \label{deltae}
\end{equation}
where $L$ is the size of the system, $|n,g\rangle$ and $E_n(g)$ are
the eigenstates and eigenvalues of $\hat H(g)$. Typically $\Delta E$
is much larger than the energy scale $E_c$ of the low-energy
excitations at criticality $g=g_c$, which is $E_c\sim L^{-z}$ with
$z=1$ for continuous transitions of Ising-like systems. This
would naturally lead to the expectation that the unitary
energy-conserving dynamics after quenching to $g_c$ is not
significantly related to the quantum critical features of the
low-energy spectrum of the critical Hamiltonian.  However, as argued
in Refs.~\cite{BDD-15,RMD-17,TIGGG-19,HPD-18,HMHPRD-20}, some
signatures may emerge as well. In particular, integrable many body
systems (such as systems that are mappable into generic noninteracting
fermionic systems) develop some peculiar discontinuities even in the
asymptotic stationary states arising from the quantum
quenches~\cite{BDD-15, RMD-17}.  On the other hand, local observables
are not expected to present singularities in nonintegrable systems
where generic quantum quenches lead to thermalization, since they are
generally smooth functions of the temperature.  Nonetheless it has
been recently argued that it is possible to recover some signals from
intermediate regimes of the quantum evolution after
quenches~\cite{HMHPRD-20}.

In this paper we elaborate on this issue, focusing on the
out-of-equilibrium dynamics arising from quantum quenches within
one-dimensional Ising-like quantum systems.  Specifically, we consider
quantum XY chains, for which we observe peculiar nonanalyticities in
the behavior of the transverse and longitudinal magnetizations, when
comparing them after quenches of the Hamiltonian parameter $g$ around the
critical point.  We also extend the analysis to finite-size
effects, showing the emergence of revival phenomena with peculiar
scaling behaviors. This somehow extends the phenomenology of quantum
revival phenomena observed in various contexts, see
e.g. Refs.~\cite{HHH-12,KLM-14,Cardy-14,JH-17,MAC-20}.  Moreover, we
discuss the impact of dissipative mechanisms arising from interactions
with an environment, and the effects of integrability-breaking
perturbations (such as next-to-nearest neighbor Ising-like
couplings). Our purpose is to understand whether, and in which
conditions, it is still possible to unveil signatures of the various
equilibrium phases and quantum transitions separating them, after
a quench protocol.

The paper is organized as follows. In Sec.~\ref{model} we present the
XY chain, the dissipation modeling based on the Lindblad master
equations, and the observables that we consider during the post-quench
time evolution. Sec.~\ref{qhdispha} is devoted to the study
of the dynamics arising from quenches starting from the disordered
phase, in the thermodynamic limit and for finite-size systems with
periodic and open boundary conditions.  In Sec.~\ref{dissipation} we
study the effects of dissipation due to the interaction with an
environment within the quadratic fermionic model obtained by a
Jordan-Wigner mapping of the Ising chain.  In Sec.~\ref{quordsta} we
discuss quenches form the ordered states breaking the $\mathbb{Z}_2$
symmetry of Ising-like systems. In Sec.~\ref{ANNNI} we address some
effects of integrability breaking, due to further lattice
interactions, such as models with next-to-nearest neighbor
couplings. Finally, in Sec.~\ref{conclusions} we summarize and draw
our conclusions.

\section{The XY model and observables}
\label{model}

The ferromagnetic quantum XY chain is defined by the Hamiltonian
\begin{equation}
  \hat H_{\rm XY} = - {J\over 2} \sum_{x=1}^L \Bigl[
    (1+\gamma) \hat \sigma^{(1)}_{x\phantom{1}} \hat \sigma^{(1)}_{x+1} +
    (1-\gamma) \hat \sigma^{(2)}_{x\phantom{2}} \hat \sigma^{(2)}_{x+1} 
    + g \, \hat \sigma^{(3)}_x \Bigr]\,,
  \label{hedefxy}
\end{equation}
where $L$ is the number of lattice sites, $\hat \sigma^{(\alpha)}_x$
denote the spin-$1/2$ Pauli matrices ($\alpha = 1,2,3$) for each
lattice site ($x=1, \ldots, L$), $J>0$ is the energy scale, $\gamma$
the Ising-like coupling anisotropy, and $g$ the strength of a uniform
transverse magnetic field.  In the following we focus on positive
values of $g$.  For any $\gamma > 0$, a continuous quantum
transition~\cite{Sachdev-book} belonging to the two-dimensional Ising
universality class occurs at the critical point $g = g_c = 1$,
separating a disordered ($g > g_c$) from an ordered ($g < g_c$)
quantum phase.  For $\gamma=1$, one recovers the quantum Ising-chain
Hamiltonian
\begin{equation}
  \hat H_{\rm Is} = - J \, 
\sum_{x=1}^L \left[ \hat \sigma^{(1)}_{x\phantom{1}} \hat \sigma^{(1)}_{x+1}
    + g \, \hat \sigma^{(3)}_x \right] \,.
  \label{hedef}
\end{equation}
In the following we consider systems with boundary conditions
respecting the $\mathbb{Z}_2$ global symmetry of the model
\begin{equation}
  \hat \sigma_{x}^{(1,2)} \rightarrow - \hat \sigma_{x}^{(1,2)} \,,
  \qquad
  \hat \sigma_{x}^{(3)} \rightarrow \hat \sigma_{x}^{(3)} \,,
  \label{z1sym}
\end{equation}
such as periodic boundary conditions (PBC), in which $\hat
\sigma^{(k)}_{L+x} = \hat \sigma^{(k)}_x$, and open boundary
conditions (OBC).  Hereafter we set $J=1$.

We want to study the dynamics arising from instantaneous quenches of
the Hamiltonian parameter $g$, starting from states $|\Psi(0)\rangle$
corresponding to the ground states associated with Hamiltonian
parameters $g_0 \neq g$. In particular, as reference states, we
consider the extreme cases of maximally disordered initial state
$|\!\! \rightarrow, \ldots, \rightarrow \rangle$, corresponding to the
ground state for $g_0\to+\infty$ (i.e., all spins aligned along the
direction $\alpha = 3$), and the completely ordered state $|\!\!
\uparrow, \ldots, \uparrow \rangle$, corresponding to one of the
degenerate lowest-energy states at $g_0=0$ in the thermodynamic limit
[i.e., all spins aligned along the direction $\alpha = 1$ for the
  Ising model (\ref{hedef})]. Note that, in the latter case, the
initial state breaks the $\mathbb{Z}_2$ symmetry of the model.

For closed systems, the dynamics after a quench is unitary and the
state at any time remains pure, as given by Eq.~\eqref{unitary}.  In
such case, to monitor the resulting out-of-equilibrium time evolution,
we consider the longitudinal and transverse local magnetizations
\begin{subequations}
  \begin{eqnarray}
  M_x(t) & = & \langle\Psi(t) |\hat \sigma^{(1)}_x|\Psi(t) \rangle \, ,
  \label{mlj}\\
  S_x(t) & = & \langle\Psi(t) |\hat \sigma^{(3)}_x|\Psi(t) \rangle \, . 
  \label{ntj}
\end{eqnarray}
\end{subequations}
In the case of boundary conditions and protocols respecting
translational invariance, such as PBC or antiperiodic boundary
conditions (ABC), the above magnetizations do not depend on the
lattice site, thus
\begin{equation}
  M(t)\equiv M_x(t),\qquad S(t)\equiv S_x(t).
  \label{mndef}
\end{equation}
In the case of OBC, giving rise to inhomogeneous dependences, we
distinguish the magnetizations at the center ($c$) of the chain and at
the boundary ($b$), i.e,
\begin{subequations}
\begin{eqnarray}
  &M_c(t)\equiv M_{L/2}(t),\qquad & S_c(t)\equiv S_{L/2}(t),
  \label{mncdef}\\
  & M_b(t)\equiv M_{1}(t),\qquad &S_b(t)\equiv S_{1}(t),
\label{mnbdef}
\end{eqnarray}
\end{subequations}
where $L$ has been assumed even, for the sake of clarity.

It is worth pointing out that the time dependence of the transverse
magnetization is related to the average work $W$ necessary to perform
the following protocol: (i)~the system is initially prepared in the
ground state associated with a Hamiltonian transverse parameter $g_0$;
(ii)~at $t=0$ one performs an instantaneous quench of the transverse
parameter to $g\neq g_0$; (iii)~then the time evolution is driven by
the Hamiltonian $\hat H(g)$; (iv)~after a time $t_f$, the transverse
parameter is instantaneously brought back to $g_0$, and the system
evolution is again driven by the initial Hamiltonian $\hat H(g_0)$.
The total average work can be obtained as
\begin{eqnarray}
  W & = & E_f - E_i \,,\label{avworkdef}\\
  E_f & = & \langle \Psi(t\ge t_f) | \hat H(g_0) | \Psi(t\ge t_f) \rangle \,,
\nonumber\\
  E_i & = & \langle \Psi(t=0) | \hat H(g_0) | \Psi(t=0) \rangle\,, \nonumber
\end{eqnarray}
where we used the fact that, after $t_f$, the average energy is
conserved along the unitary evolution.  Straightforward calculations
lead to the expression
\begin{equation}
  W = (g-g_0) \sum_{x=1}^L \big[ S_x(t_f) - S_x(0) \big] \,.
  \label{avwork}
\end{equation}
Thus, for translationally invariant systems, we obtain
\begin{equation}
  W/L = (g-g_0) \,[ S(t_f) - S(0) \big] \,.
  \label{avwork2}
\end{equation}

If the system is in contact with some environment, the unitary
evolution (\ref{unitary}) is not valid anymore, because the time
dependence is also determined by nonunitary drivings arising from the
interaction with the environment.  The state of the system at a given
time $t$ is represented by a density matrix $\rho(t)$ whose time
dependence can be reasonably described by the Lindblad master
equation~\cite{BP-book}
\begin{equation}
  {\partial\rho\over \partial t} = -i \big[ \hat H(g),\rho \big] 
+ u \,{\mathbb D}[\rho]\,.
  \label{lindblaseq}
\end{equation}
The first term in the right-hand side provides the coherent driving,
while the second term accounts for the coupling to the environment,
characterized by a global coupling constant $u>0$.  In our quench
protocol, the initial state can be thus written as $\rho(t=0) =
|\Phi_{g_0}\rangle \langle \Phi_{g_0}|$
The instantaneous longitudinal and transverse magnetizations now read
\begin{equation}
  M_x(t) = {\rm Tr} \big[ \hat \sigma^{(1)}_x \rho(t) \big] \,, \qquad
  S_x(t) = {\rm Tr} \big[ \hat \sigma^{(3)}_x \rho(t) \big] \,.
  \label{mtt_mlt_diss}
\end{equation}
Analogous definitions as those in
Eqs.~\eqref{mndef},~\eqref{mncdef},~\eqref{mnbdef}, can be adopted
here as well, depending on the choice of the boundary conditions.

\section{Quenches starting from the disordered phase}
\label{qhdispha}

We first focus on the unitary dynamic behavior of closed systems which
arises from instantaneous changes of the transverse parameter $g$,
starting from the ground state associated with an initial value $g_0 >
1$ in the disordered phase. By symmetry, the longitudinal
magnetization remains zero at any time after the quench, $M_x(t)=0$,
while the transverse magnetization $S_x(t)$ presents a nontrivial time
dependence~\cite{Niemeijer-67, BMD-70}. We initially discuss the case
of PBC for which $S(t)\equiv S_x(t)$ independently of the lattice
site, and then that of OBC, for which we present results for both the
central and boundary local transverse magnetizations, $S_c$ and $S_b$
respectively.

\begin{figure}[!b]
  \includegraphics[width=0.95\columnwidth]{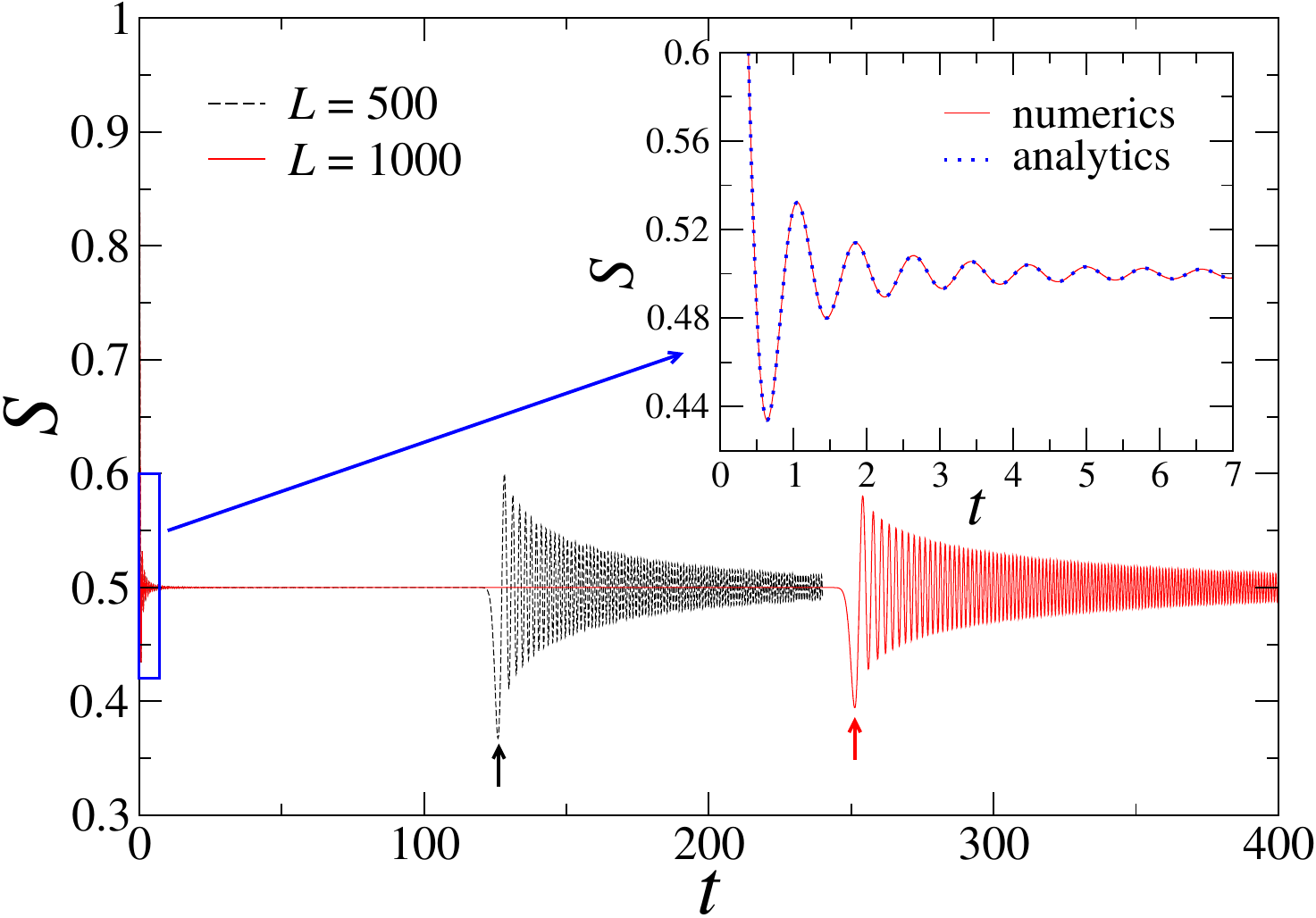}
  \caption{Transverse magnetization $S(t)$ after a quench in the
    quantum Ising chain with PBC, starting from a completely
    disordered state ($g_0 = + \infty$), to $g=1$ (critical point),
    for $L=500$ up to $t=240$ (dashed black line), and for $L=1000$ up
    to $t=400$ (continuous red line).  At relatively small times,
    $t<100$, the data sets superimpose, corresponding to the
    thermodynamic limit, while after some time which depends on the
    size, in particular after $t \approx L/(2v_m)$ with $v_m=2 \, {\rm
      Min}[g,1]=2$ (indicated by the arrows), peculiar oscillating
    behaviors appear, corresponding to finite-size revivals.  The
    inset shows a magnification at small times (blue box), where
    numerical data have been superimposed with the analytic results in
    the thermodynamic limit (dotted blue line) obtained in
    Ref.~\cite{BMD-70}.}
    \label{fig:Kitaev_t_g1}
\end{figure}

A representative example is shown in Fig.~\ref{fig:Kitaev_t_g1}, for
an instantaneous quench to $g_c$, starting from the fully polarized
state along the transverse direction, which corresponds to the ground
state $|\Phi_{g_0 \to \infty}\rangle$.  The considered system is
subject to PBC.  The time evolutions of the transverse magnetization
for the sizes $L=500$ and $L=1000$ appear identical up to $t\lesssim
100$.  Within this interval the curves rapidly converge toward the
value $S(\infty)=0.5$ (see inset of Fig.~\ref{fig:Kitaev_t_g1}), in a
time interval $t\lesssim 10$, and then appear constant up to $t\approx
100$.  Since data for $L=500$ and $L=1000$ match, this behavior should
correspond to the time dependence in the thermodynamic limit
$L\to\infty$. However, after such a relatively large interval, where
the systems have apparently reached their stationary behavior, the
time dependence becomes again nontrivial.  It shows peculiar
structures at later times depending on the size, starting from
$t\approx 125$ for $L=500$ and $t\approx 250$ for $L=1000$, thus at
larger and larger times with increasing the size. These are
characterized by an oscillatory behavior enveloped by a smooth
decreasing function, and are clearly finite-size effects.  In the
following we characterize the two regimes, identifying their main
features.  We conclude this brief summary by noticing that
Fig.~\ref{fig:Kitaev_t_g1} reports the results for a chain with PBC,
but analogous qualitative considerations apply to different boundary
conditions as well (see, e.g., Sec.~\ref{fssdis_obc} for the case with
OBC).

\subsection{Time dependence in the thermodynamic limit}
\label{thtrma}

The time dependence of the transverse magnetization in the
thermodynamic limit was analytically computed long time ago by
Niemeijer~\cite{Niemeijer-67} and by Barouch, McCoy, and
Dresden~\cite{BMD-70}.  The infinite-size limit of the transverse
magnetization,
\begin{equation}
\Sigma(t,g_0,g) \equiv 
S(t,g_0,g,L\to\infty) 
\,, \label{asymt}
\end{equation}
can be written as a sum of an asymptotic time-independent term and a
time dependent term vanishing in the large-time limit, i.e.
\begin{equation}
 \Sigma(t,g_0,g) = F(g_0,g) + F_t(t,g_0,g) \,.
  \label{faf}
\end{equation}
Setting 
\begin{equation}
  \Lambda(k,g) = \sqrt{[g - \cos(k)]^2 + \gamma^2 \sin(k)^2} \,,
  \label{lagk}
\end{equation}
Ref.~\cite{BMD-70} reports
\begin{subequations}
\begin{eqnarray}
  F(g_0,g) & = & \int_0^\pi {dk\over \pi} \, 
\frac{g-\cos(k)}{\Lambda(k,g_0)
    \Lambda(k,g)^2} \times \label{fagg0}\\ 
  & \times &
  \left\{[g_0-\cos(k)][g-\cos(k)] +\gamma^2
  \sin(k)^2\right\}\,, \nonumber
\end{eqnarray}
and
\begin{equation}
  F_t(t,g_0,g) = \int_0^\pi {dk\over \pi} \,
  {\gamma^2 (g_0-g) \sin(k)^2 \cos[4\Lambda(k,g)t] \over \Lambda(k,g_0)
    \Lambda(k,g)^2} \label{ftgg0} \,.
\end{equation}
\end{subequations}
In particular, for $g_0\to\infty$ these expressions simplify into
\begin{subequations}
\begin{eqnarray}
  F(\infty,g) & = & \int_0^\pi {dk\over \pi} 
\; {[g-\cos(k)]^2\over \Lambda(k,g)^2} \,,
  \label{fagg0inf} \\
  F_t(t,\infty,g) & = & \int_0^\pi {dk\over \pi} \;
  {\gamma^2 \sin(k)^2 \cos[4\Lambda(k,g)t] \over \Lambda(k,g)^2} \,.\qquad
  \label{ftgg0inf}
\end{eqnarray}
\end{subequations}

\begin{figure}%[!t]
  \includegraphics[width=0.95\columnwidth]{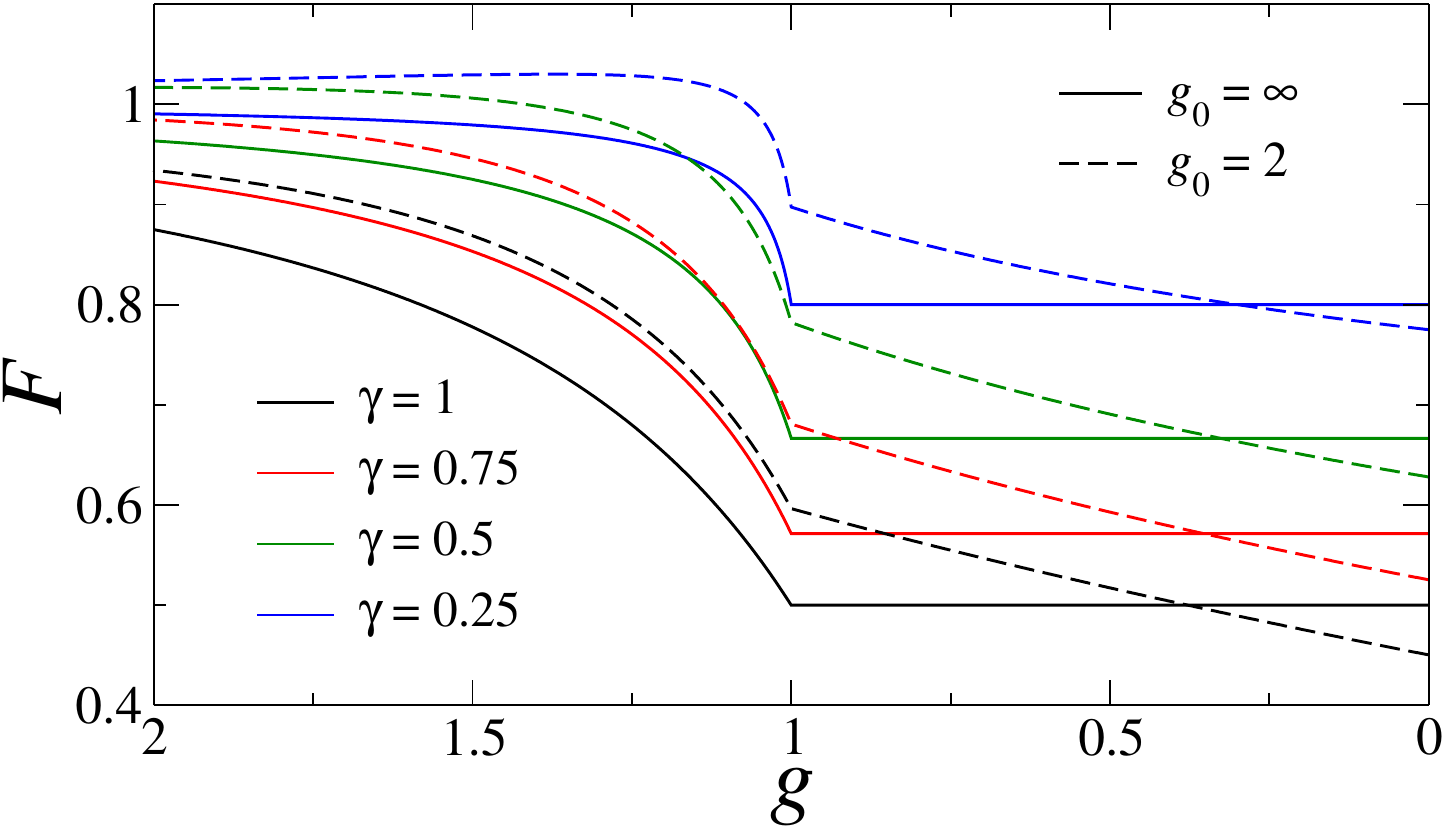}
  \caption{The function $F(g_0,g)$ of Eq.~\eqref{fagg0} representing
    the large-time limit of the transverse magnetization in the
    thermodynamic limit~\cite{BMD-70}, as a function of $g$, for
    various values of $\gamma$ (different colors, see legend), and
    $g_0=\infty$ (continuous lines) or $g_0=2$ (dashed lines).  The
    curves display a singular behavior, i.e. a discontinuity in the
    derivative, at $g=g_c=1$.}
    \label{Mt_asynt_an}
\end{figure}

Let us first focus on the large-time limit~\eqref{fagg0} of
$\Sigma(t)$.  As shown by the plots in Fig.~\ref{Mt_asynt_an}, the
function $F(g_0,g)$ presents a nonanalytic behavior in correspondence
of the critical point $g_c = 1$. Indeed, we have that
\begin{equation}
  \lim_{g\to g_c^+} {\partial F(g_0,g)\over \partial g}
  - \lim_{g\to g_c^-} {\partial F(g_0,g)\over \partial g}  = \gamma^{-1}\,.
  \label{singdermt}
\end{equation}
Note that such a discontinuity is independent of $g_0>1$.  In
particular, for $\gamma=1$ and $g_0\to\infty$, the behavior around
$g_c$ turns out to be
\begin{equation}
  F(\infty,g) \!= \!\left\{ \begin{array}{ll}
    \!{1\over 2} + (g\!-\!1) + O[(g\!-\!1)^2] \; & {\rm for} 
\; g > 1 \,, \vspace*{2mm}\\
    \!{1\over 2}                      \; & {\rm for} \; g\leq 1 \,.
  \end{array} \right. \label{fag1}
\end{equation}
An analogous discontinuity has been reported in Ref.~\cite{BDD-15} for
the asymptotic behavior of the energy density after a quench, and also
in generic noninteracting fermionic systems~\cite{RMD-17}.

\begin{figure}%[!t]
  \includegraphics[width=0.95\columnwidth]{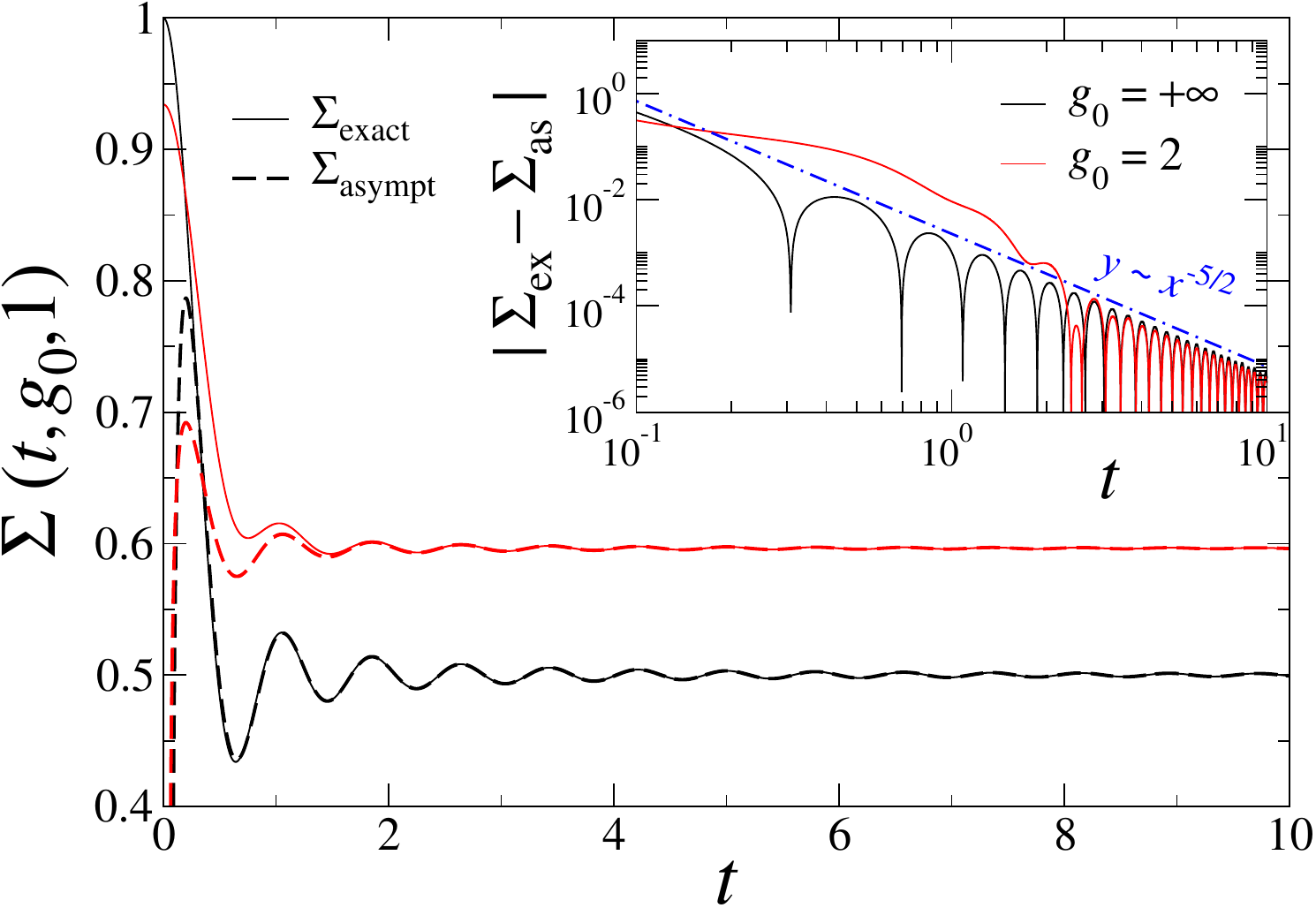}
  \caption{Time behavior of the transverse magnetization
    $\Sigma(t,g_0,g)$ in the infinite-size limit, for a quench 
    in the quantum Ising chain from
    two different values of $g_0$ (see legend), to the critical point
    $g=1$. Continuous lines denote the exact result obtained using
    Eq.~\eqref{ftgg0}, while dashed lines report the approximate
    result using Eq.~\eqref{fr1infty}.  The inset shows absolute
    discrepancies between exact and approximate results; the
    dot-dashed blue line denotes a $t^{-5/2}$ behavior and is plotted
    to guide the eye.}
    \label{Mt_asynt_an_g1}
\end{figure}

Even the asymptotic approach to the large-time limit is singular at
$g_c$.  Indeed for $g\neq 1$ it is given by~\cite{BMD-70,CEF-12-2}
\begin{eqnarray}
  &&F_t(t,g_0,g) = {g_0 - g\over 2^{7/2} \sqrt{\pi} g^{3/2}} t^{-3/2}
  \times \nonumber\\
  &&\qquad\times\left[ {\sin[4(g+1) t - \pi/4]\over (g_0+1) \sqrt{g+1}}
    -{\sin(4|g-1| t + \pi/4)\over |g_0-1| \sqrt{|g-1|}} \right]
  \nonumber\\
  && \qquad + \;O(t^{-5/2})\,,
  \label{ftgg0_t}
\end{eqnarray}
while at $g=1$ it turns out to be given by
\begin{equation}
  F_t(t,g_0,1) = {(g_0-1) \sin(8 t - \pi/4) \over 16 (g_0+1) \sqrt{\pi}} \,
  t^{-3/2} + O(t^{-5/2})\,.
  \label{fr1infty}
\end{equation}
Note that Eq.~\eqref{fr1infty} is simply obtained by dropping the
second divergent term within the parenthesis in Eq.~\eqref{ftgg0_t}.
Numerical evidence of the validity of Eq.~\eqref{fr1infty}, for
sufficiently long times, is provided by the data shown in
Fig.~\ref{Mt_asynt_an_g1}. There we report a direct comparison between
the transverse magnetization $\Sigma(t,g_0,g)$ of Eq.~\eqref{faf},
calculated either with the exact formula of Eq.~\eqref{ftgg0} or with
the approximate one of Eq.~\eqref{fr1infty}.  The agreement between
the two results emerges from the analysis of the absolute discrepancy
between them, explicitly shown in the inset, where corrections of
$O(t^{-5/2})$ evidently appear (see dot-dashed blue line).

We finally recall that the thermodynamic limit, approached by taking
the large-$L$ limit at fixed time, is expected to be independent of
the boundary conditions.  This is confirmed by our numerical results.
However, as we shall see below, the finite-size effects are instead
dependent on the boundary conditions. In the following we address the
cases of PBC and OBC separately.

\subsection{Finite-size effects with PBC}
\label{fssdis_pbc}

We now discuss the emergence of finite-size effects mostly related to
revival phenomena, which are characterized by definite power-law
scaling behaviors.  Analyses of revival finite-size effects
have been recently discussed under various perspectives
in Refs.~\cite{RI-11,IR-11,BRI-12,MAC-20}, considering also
the behavior of the entanglement properties.
Here we focus on the quantum Ising chain~\eqref{hedef} with PBC.

Before presenting our results, it is instructive to realize that, for
quenches starting from ground states for $g_0>g_c$, the dynamic
problem can be exactly mapped into that of a fermionic quadratic model
with antiperiodic boundary conditions (ABC), see
e.g. Ref.~\cite{CPV-14}.  This is essentially due to the fact that the
initial quantum state $|\Phi_{g_0>g_c}\rangle$ has a definite parity,
and therefore only fermionic states with ABC are involved during the
evolution.
Therefore, for these types of quenches, we may consider the equivalent
Kitaev quantum wire defined by the Hamiltonian~\cite{Kitaev-01}
\begin{equation}
 \hat H_{\rm K} = - \sum_{x=1}^L \big( \hat c_x^\dagger \hat
  c_{x+1} + \gamma \, \hat c_x^\dagger \hat c_{x+1}^\dagger+{\rm h.c.}
  \big) - 2 g \sum_{x=1}^L \hat n_x \,,
  \label{kitaev2}
\end{equation}
where $\hat c_x$ is the fermionic annihilation operator on the $x$th
site of the chain and $\hat n_x\equiv \hat c_x^\dagger \hat c_x$ the
density operator.  The quench problem considered here can be matched
by taking ABC, $\hat c_{L+x} = - \hat c_x$ (we suppose $L$ even, for
convenience). The straightforward diagonalization of the quadratic
fermionic model~\eqref{kitaev2} allows one to obtain results for very
large sizes, up to $L = O(10^4)$, requiring computational resources
which increase only linearly with $L$.  For details see, e.g.,
Refs.~\cite{CPV-14,RV-20}. In the following we set $\gamma=1$,
corresponding to the Ising chain~\eqref{hedef}.

\begin{figure}%[!t]
\includegraphics[width=0.95\columnwidth]{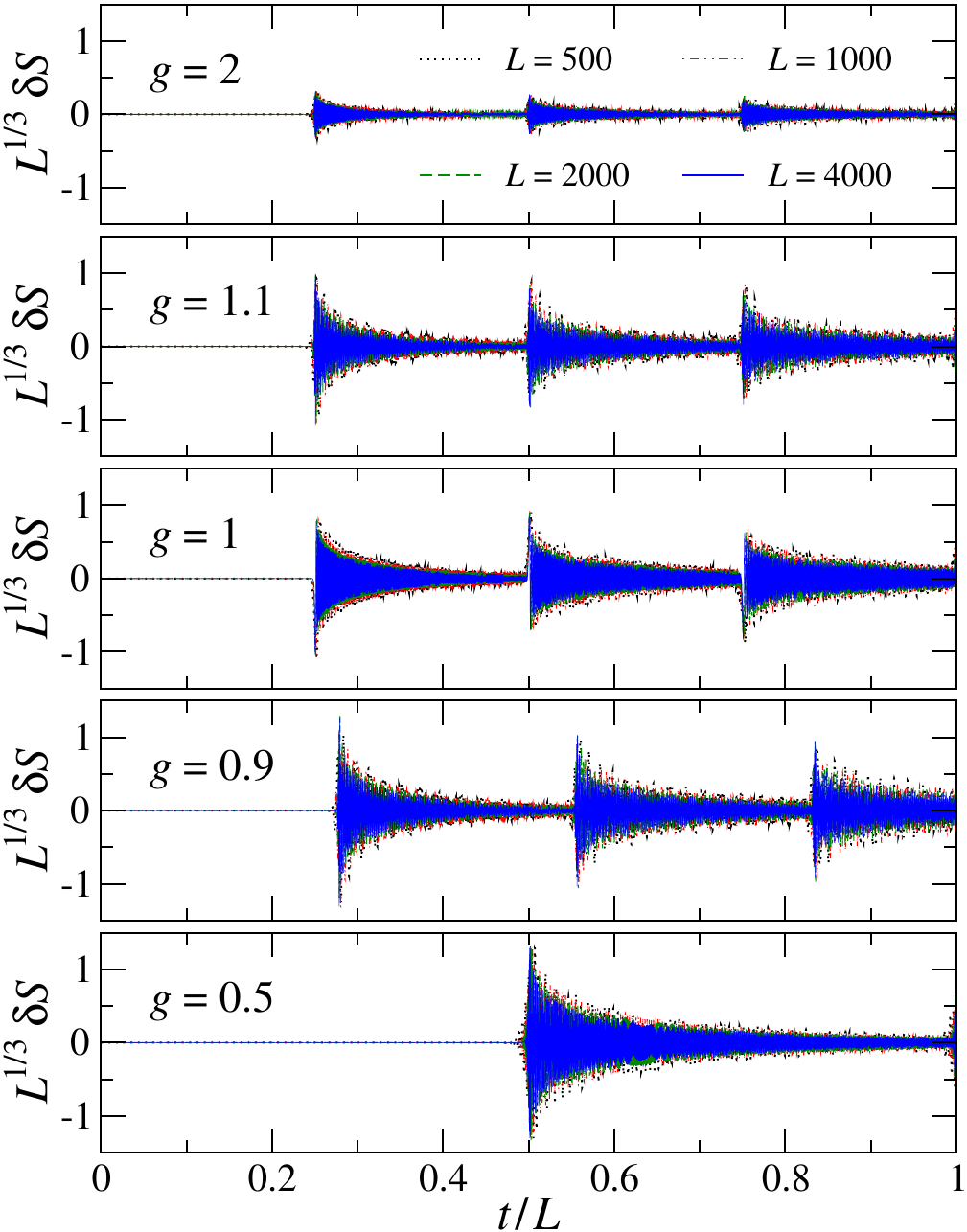}
\caption{Finite-size features of the temporal behavior of the
  transverse magnetization after a quench from $g_0=+\infty$.  We plot
  $L^{1/3} \delta S$ versus the rescaled time $t_L\equiv t/L$, where
  $\delta S$ is defined in Eq.~\eqref{submat}.  Different panels are
  for $g=2, \, 1.1, \,1, \, 0.9, \,0.5$.  Colored curves stand for
  various system sizes, as indicated in the legend, and nicely support
  the behavior put forward in Eq.~\eqref{asyd}.}
\label{fig:Rescal1_full_g}
\end{figure}

\begin{figure}%[!t]
  \includegraphics[width=0.95\columnwidth]{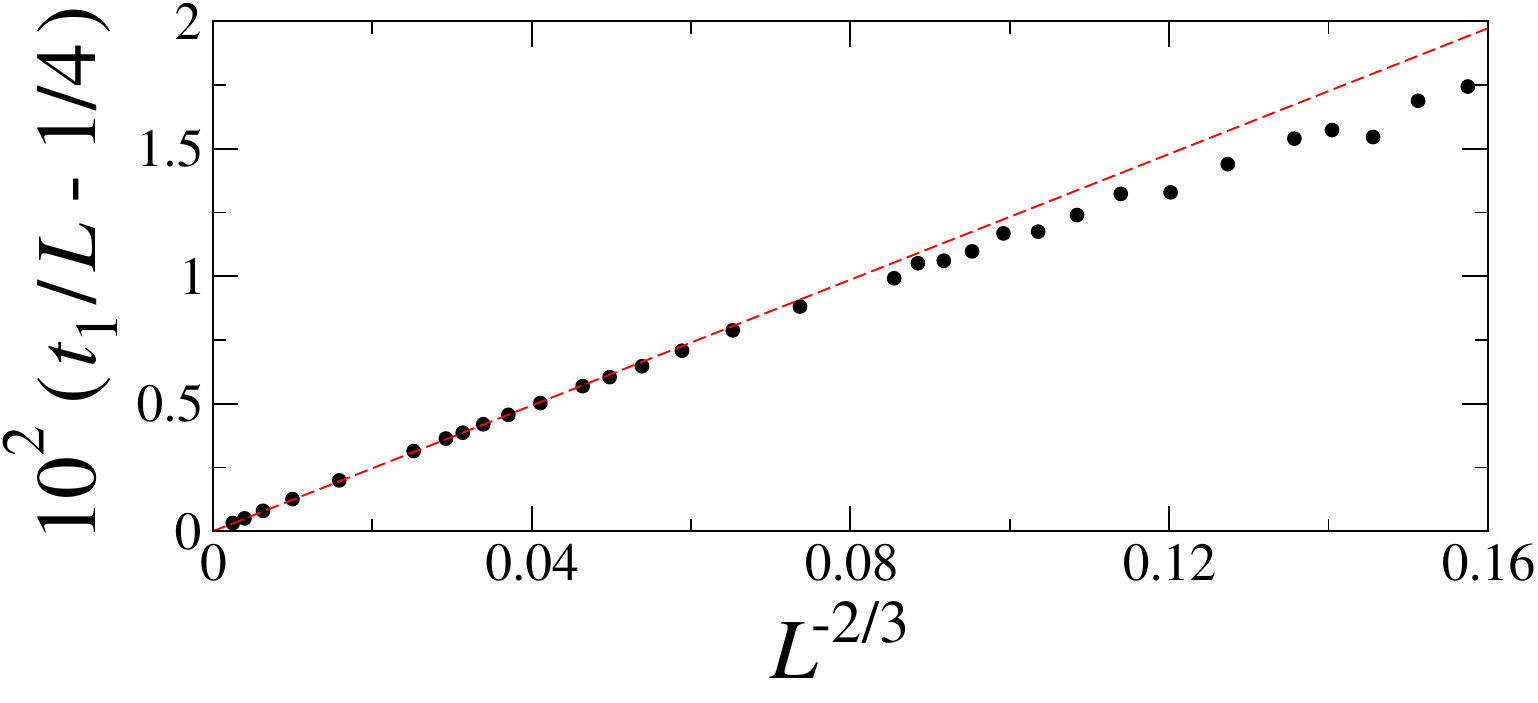}
  \includegraphics[width=0.95\columnwidth]{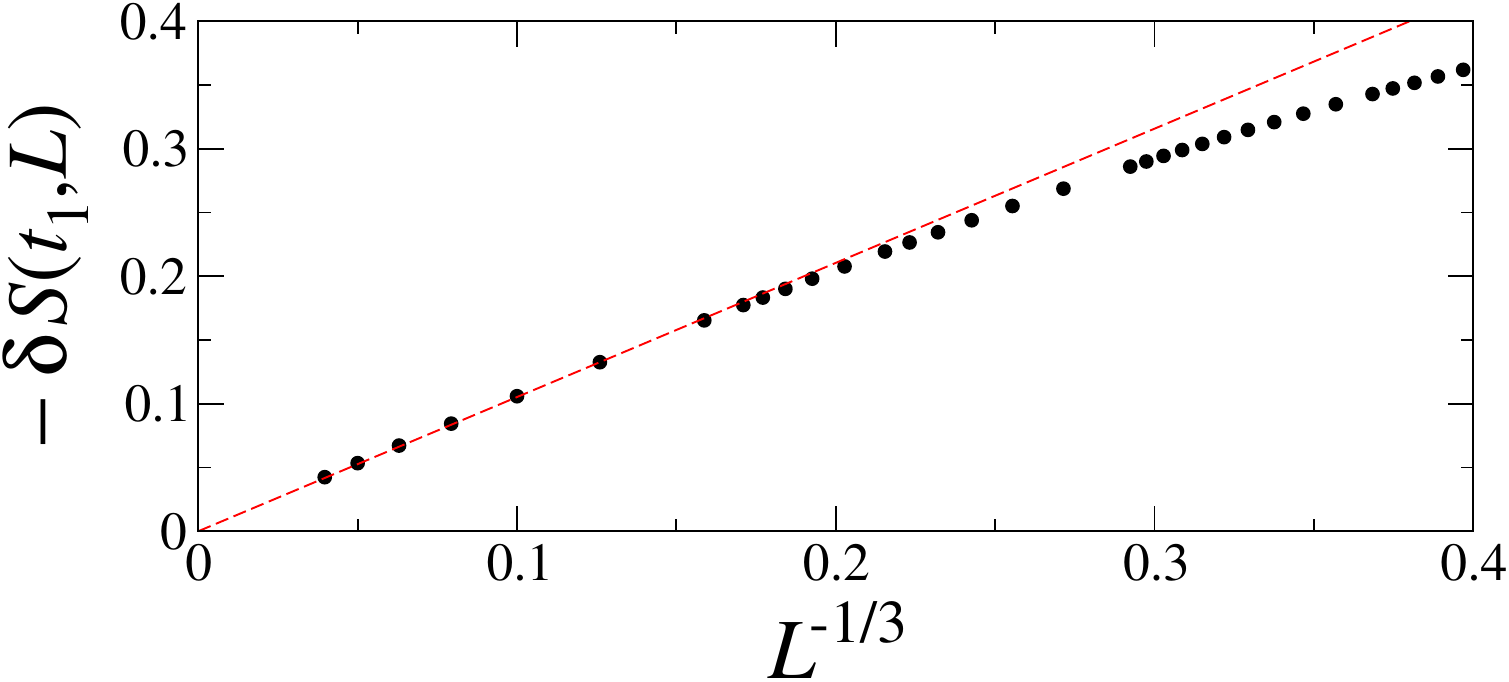}
  \caption{Scaling behavior of the position and height of the first
    dip (located at $t_1\approx L/4$, corresponding to $t_{L,1}\equiv
    t_1/L \approx 1/4$) of the transverse magnetization for quenches
    to the critical point $g=1$, starting from the fully disordered
    state $|\Phi_{\infty}\rangle$ (see arrows in
    Fig.~\ref{fig:Kitaev_t_g1}).  We simulated systems with up to
    $L=10^4$ sites.  Upper panel: data for $t_{L,1} - 1/4$
    vs.~$L^{-2/3}$, confirming the behavior predicted by
    Eq.~\eqref{taupeak}; the dashed red line shows a linear fit for
    $L\ge 100$.  Lower panel: data for $\delta S(t_1,L)$
    vs.~$L^{-1/3}$; the dashed red line shows a linear fit for $L\ge
    250$.}
    \label{fig:Peak_XY_Kitaev_g1}
\end{figure}

Some results for quenches from $g_0\to \infty$ to various values of
$g$ are shown in Fig.~\ref{fig:Rescal1_full_g}, where we report the
subtracted transverse magnetization
\begin{equation}
  \delta S(t,g_0,g,L) \equiv S(t,g_0,g,L)  - \Sigma(t,g_0,g)\,,
  \label{submat}
\end{equation}
$\Sigma(t,g_0,g)$ being the infinite-size limit given by
Eq.~\eqref{asymt}.  The numerical results show the following behavior
\begin{equation}
  \delta S(t,L) = L^{-a} f_e(t_L) f_o(t,L) + O(L^{-1})\,,  
  \label{asyd}
\end{equation}
where 
\begin{equation}
t_L \equiv t/L \,,\qquad a = 1/3 \,
  \label{kappa_tldef}
\end{equation}
(as we shall see, the accuracy of the estimate of exponent $a$ is very
high), $f_o(t,L)$ is a rapidly oscillating function around zero
depending on both $t$ and $L$, while the envelope function $f_e$ is a
(non-oscillating) function of $t_L$ with discontinuities located at
$t_L=t_{L,k}$ [for simplicity, here we have omitted the dependence on
  $g_0$ and $g$ in Eq.~(\ref{asyd})].
  
The discontinuities of $f_e$ are essentially related to revival
phenomena and appear at times
\begin{equation}
  t_{L,k} \equiv  {k\over 2v_m}\,,\qquad v_m = 2 \, {\rm Min} [g,1]\,,
  \label{tk}
\end{equation}
for $k = 1,2, \ldots$, where $v_m$ is the maximum velocity of the
quasi-particle modes~\cite{LR-72,CC-05,CEF-12-1}.  The amplitude of
such discontinuities generally tends to decrease with increasing $k$,
as it can be seen from the various panels of
Fig.~\ref{fig:Rescal1_full_g}.  We point out that the scaling
behavior of the revival times $t_{L,k}$, and their connection
with the maximum velocity of the quasi-particle modes, were already
put forward by other studies of revival phenomena in finite-size
systems, see e.g. Refs.~\cite{RI-11,BRI-12,MAC-20}.

Our numerical results for $g=g_c=1$ show that the first
sharp dip is asymptotically located at
\begin{equation}
  t_{L,1} = 1/4 + O(L^{-2/3})\,,
  \label{taupeak}
\end{equation}
as reported in the upper panel of Fig.~\ref{fig:Peak_XY_Kitaev_g1}.
The relative accuracy achieved by our numerical results on the
asymptotic location of the peak is very high, within $O(10^{-5})$, so
that we can safely conclude that its value is 1/4 with great
accuracy. This can be related to the interference between the signals
traveling in the opposite direction with velocity $v_{m} =
2$,~\cite{KLM-14} taking a time $t = L/(2 v_m) = L/4$ to approach each
other.  Note that the $O(L^{-2/3})$ corrections to the value of $t_L$
at the dip arise from the $O(L^{-1})$ corrections in the
formula~\eqref{asyd}.  We mention that, once fixed the asymptotic
value $t_{L,1}=1/4$, the relative accuracy of the estimate of the
exponent of the power-law correction in Eq.~(\ref{taupeak}) is a few
per mille.  The lower panel of Fig.~\ref{fig:Peak_XY_Kitaev_g1} shows
the large-$L$ scaling of the transverse magnetization at the dip,
clearly demonstrating the power-law asymptotic behavior $L^{-a}$ with
$a=1/3$, which is approached with $O(L^{-2/3})$ corrections, as
implied by Eq.~\eqref{asyd}. The relative accuracy on the estimate of
$a$ turns out to be safely better than $10^{-4}$.  Therefore, somehow
biased by the expectation that the exponent $a$ should be a simple
fraction, we assume $a=1/3$ in the following.
%Analogous results are obtained for any finite value of $g_0>1$.

\begin{figure}%[!t]
\includegraphics[width=0.95\columnwidth]{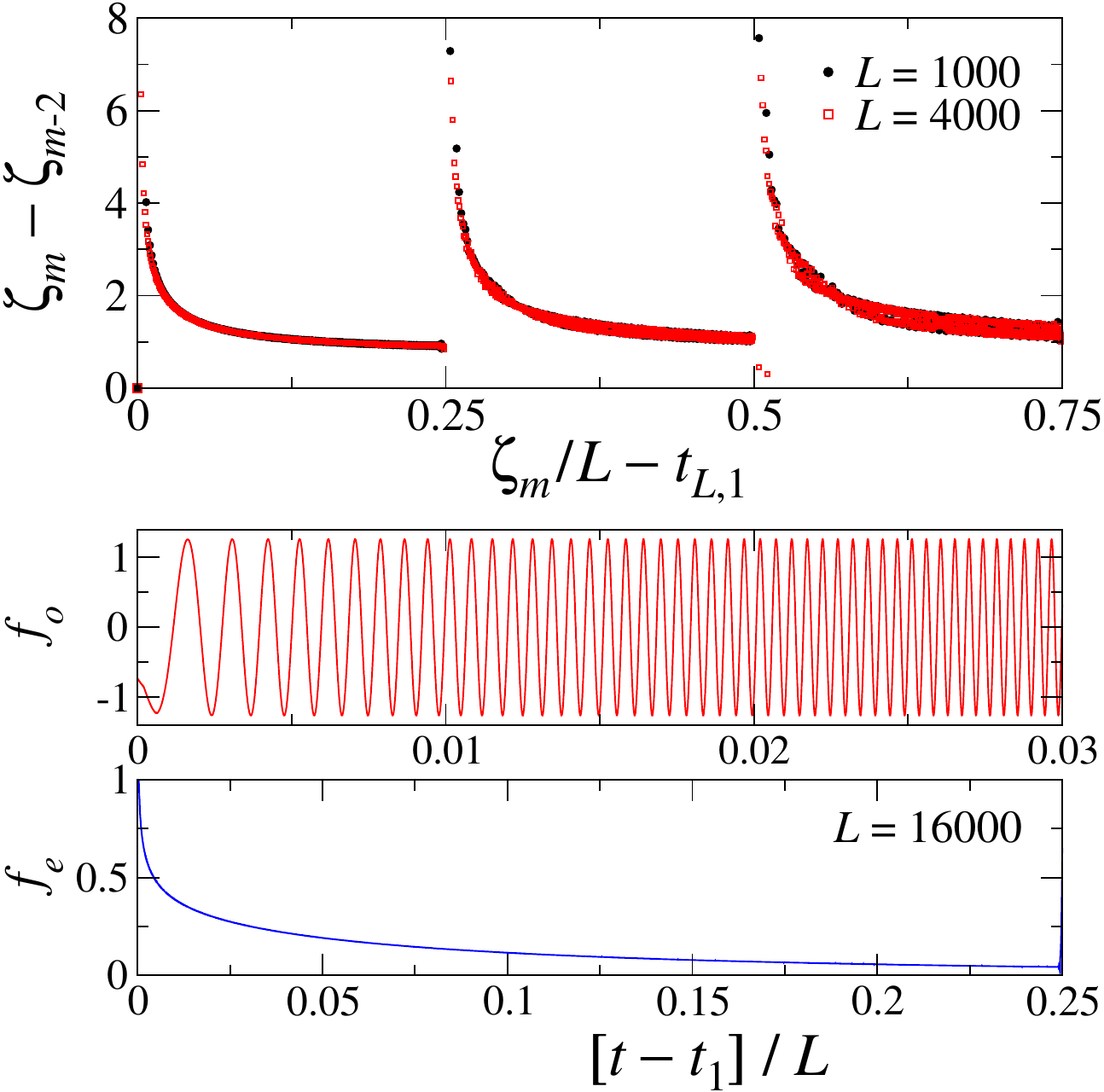}
\caption{Numerical analysis of the oscillations appearing in the time
  behavior of $\delta S(t,L)$, after a quench from the completely
  disordered state to $g_c$.  Top panel: temporal interval
  $\zeta_m-\zeta_{m-2}$ between the times at which the $m$th zero and
  the $(m-2)$th zero occur, for two different sizes $L$.  This is
  plotted as a function of the rescaled time $\zeta_m/L -t_{L,1}$.
  Central panel: time behavior of the oscillating function $f_o$,
  numerically inferred by taking the ratio between $L^{1/3} \delta
  S(t,L)$ at a given size $L$ ($L=4000$ in the figure) and the
  interpolation of the maxima of the corresponding curve for the
  largest available data set at $L=16000$. The latter also gives an
  estimate of the envelope function $f_e$, reported in the bottom
  panel.}
\label{fig:Mt_Oscill}
\end{figure}

The zeroes of $\delta S$, cf.~Eq.~(\ref{asyd}), are essentially
controlled by the oscillating function $f_o$. In
Fig.~\ref{fig:Mt_Oscill} we show some analyses of the results for
quenches to $g=g_c=1$.  The top panel displays numerical results for
the zeroes $\zeta_m$ of the function $\delta S$.  The difference
$\zeta_m-\zeta_{m-2}$, which would correspond to the period in the
case of periodic functions, turns out to be a function of $t_L\equiv
t/L$, with singularities at $t_{L,k}$.  The bottom panel shows the
envelop function $f_e$ as obtained by interpolating the maxima of
$L^{1/3}\delta S$ for the largest available lattice size $L=16000$,
while the intermediate panel shows the resulting oscillatory function
$f_o$.  The above results globally support the ansatz~\eqref{asyd}.

\begin{figure}%[!t]
\includegraphics[width=0.95\columnwidth]{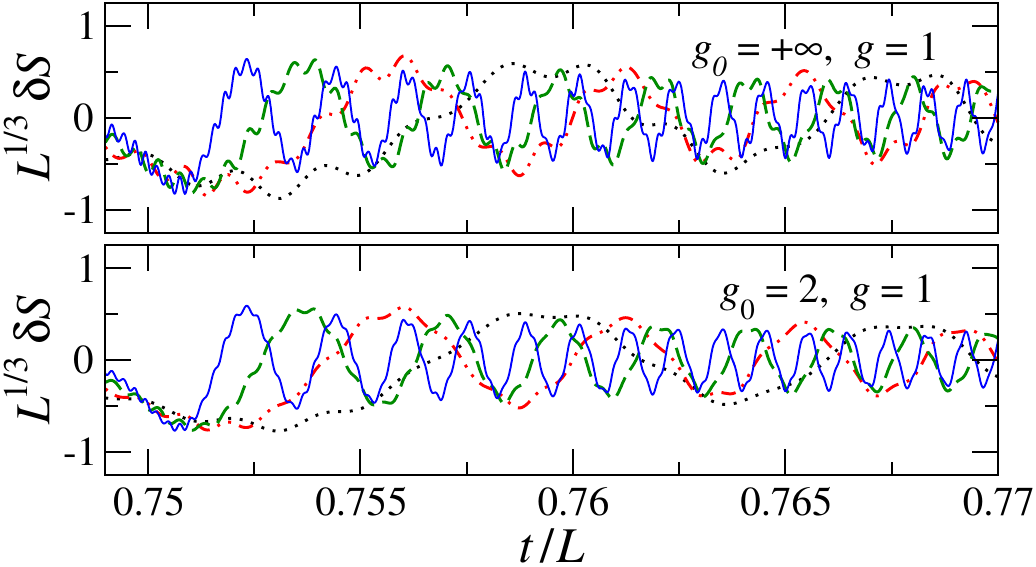}
\caption{Finite-size features of the temporal behavior of the
  transverse magnetization after a quench to the critical point $g=1$,
  starting from either $g_0=+\infty$ (upper panel) or $g_0=2$ (lower
  panel).  We plot curves for $\delta S(t,g_0,g,L)$ as defined in
  Eq.~\eqref{submat}, zooming around the third revival [see
    Eq.~\eqref{tk}].  Note that the upper panel displays the same data
  of the central panel of Fig.~\ref{fig:Rescal1_full_g}. The color
  code is the same as in that figure.}
\label{fig:Rescal1_g02}
\end{figure}

We note that the scaling behavior~\eqref{asyd} is generally observed
for any value of $g$, as confirmed by our numerical simulations (not
shown here). Of course, the functions $f_e$ and $f_o$ will depend on
$g$, but their structure looks similar when varying it. Therefore,
this shows that the main features of the finite-size effects are not
related to the existence of quantum criticality at $g_c$.
We also mention that analogous results, reproduced by Eq.~\eqref{asyd}
with the appropriate dependence on $g_0$ and $g$, are obtained when
starting from ground states corresponding to finite values of $g_0>1$.
Indeed, comparing the two panels of Fig.~\ref{fig:Rescal1_g02},
obtained for quenches starting from either $g_0=+\infty$ or $g_0=2$,
one can recognize the same scaling behavior for the curves
corresponding to various system sizes.  Note that times have been
zoomed around $t/L \approx 0.75$, in order to highlight the third
revival phenomena appearing at time $t_3$ [see Eq.~\eqref{tk}]. A
close look at the two panels reveals that, although very similar
patterns can be seen for the two cases, fast oscillating wiggles are
slightly reduced for $g_0=2$.  Such differences are even less visible
for the former two revivals (not shown).

We finally mention that finite-size revivals in the entanglement
entropy were investigated in Ref.~\cite{MAC-20}, showing that
they are also characterized by dips in the block entanglement entropy.

\subsection{Finite-size effects with OBC}
\label{fssdis_obc}

A natural arising issue concerns the dependence on the boundary
conditions of the main features observed in the previous sections.  To
answer this question, we have considered quench protocols applied to
systems with OBC. Here we report results starting from the fully
disordered state $|\Phi_{+\infty}\rangle$, focusing again on the
transverse magnetization.  However, since OBC breaks the translational
invariance, we will separately consider the central and boundary local
transverse magnetizations, cf.~Eqs.~\eqref{mncdef} and~\eqref{mnbdef},
as representative observables.  Technically, we can still take
advantage of the mapping into the Kitaev quantum wire, where ABC are
now replaced by OBC. However, in this latter case, going to Fourier
space is no longer helpful to diagonalize the Hamiltonian $\hat H_K$
and one has to perform a $2L$-dimensional Bogoliubov rotation. The
required computational resources keep growing polynomially with $L$,
thus preserving the ability to simulate sizes comparable to those for
translationally invariant systems.

The central local transverse magnetization is expected to have the
same large-$L$ limit discussed for systems with PBC.  We have also
checked it numerically.  Thus, to study finite-size effects, we again
subtract the asymptotic large-$L$ limit $\Sigma$ given by
Eq.~\eqref{faf}, i.e.
\begin{equation}
  \delta S_c(t,g_0,g,L)   = S_c(t,g_0,g,L) - \Sigma(t,g_0,g)\,,
  \label{deltasc}
\end{equation}
with $S_c$ defined as in Eq.~(\ref{mncdef}).  The transverse
magnetization $S_b$ at the boundary, cf.~Eq.~(\ref{mnbdef}), is
expected to behave differently, even in the large-$L$ limit, and its
time dependence in the thermodynamic limit is not known.

\begin{figure}%[!t]
\includegraphics[width=0.95\columnwidth]{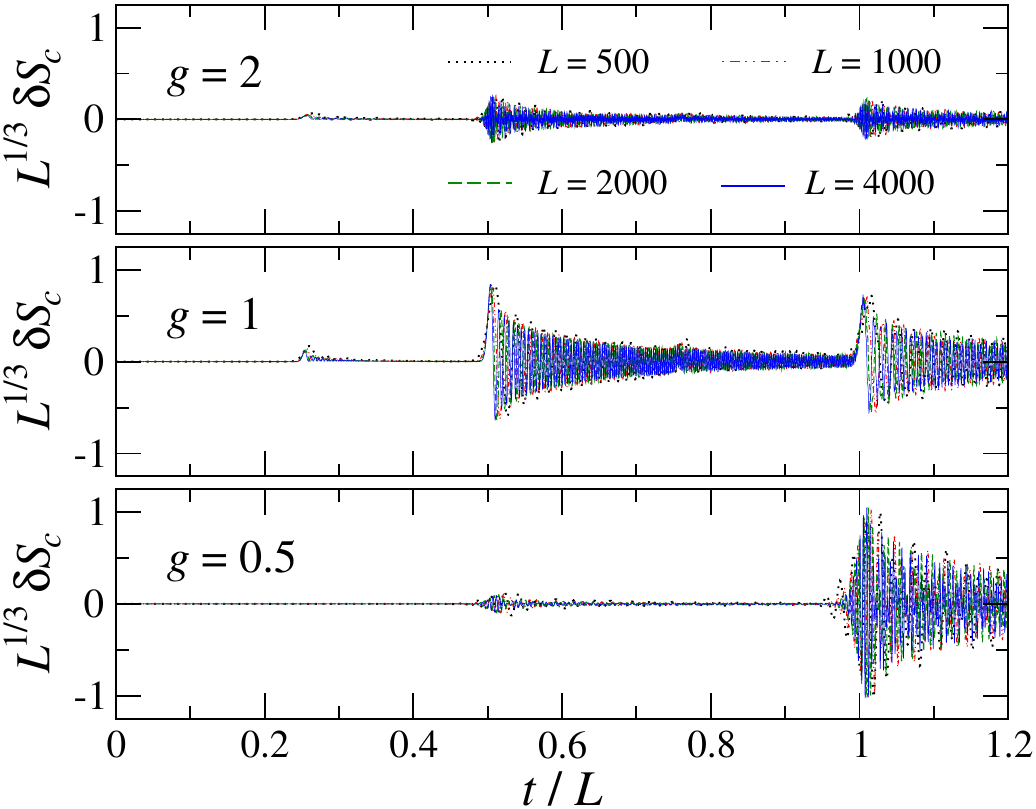}
\caption{Finite-size features of the temporal behavior of the central
  transverse magnetization, with OBC, after a quench from
  $g_0=+\infty$ to $g=2$ (upper), $g=1$ (central), and $g=0.5$ (lower
  panel).  We show curves for $L^{1/3} \delta S_c$ versus the rescaled
  time $t_L\equiv t/L$, where $\delta S_c(t,L)$ is defined in
  Eq.~\eqref{deltasc}.}
\label{fig:MtOBC_rescal}
\end{figure}

Fig.~\ref{fig:MtOBC_rescal} reports the central transverse
magnetization at the central site of the chain, when the time evolution
is controlled by the Hamiltonian at $g=2$, $g=1$, and $g=0.5$,
starting from the disordered state $|\Phi_{+\infty}\rangle$. 
The finite-size behavior of $S_c$ is qualitatively similar to
that observed for systems with PBC.  Indeed, the large-$L$ time
dependence of the subtracted central transverse magnetization $\delta
S_c$ turns out to be well described by
\begin{equation}
  \delta S_c(t,L) = L^{-a} f_e(t_L) f_o(t,L) + O(L^{-1})
  \label{asydsc}
\end{equation}
with $a = 1/3$, analogously to the PBC case [cf.~Eq.~\eqref{asyd}].
We also note that, apart from rapid oscillations encoded by the
function $f_o(t,L)$, smooth envelope structures appear, associated
with a rescaled time $t_L=t/L$. Again peculiar discontinuities emerge
in proximity of the values $t_{L,k} = k L/(2 v_m)$, for $k=1,2,\ldots$
The main difference with the PBC case is that at $t_{L,1}$ only
oscillations with tiny amplitudes emerge.  More in general, the main
structures are observed at times $t_{L,2j}$ ($j=1,2,\ldots$); small
bumps also occur at any time $t_{L,2j-1}$, although they are barely
visible for $j \geq 2$.

\begin{figure}%[!t]
\includegraphics[width=0.95\columnwidth]{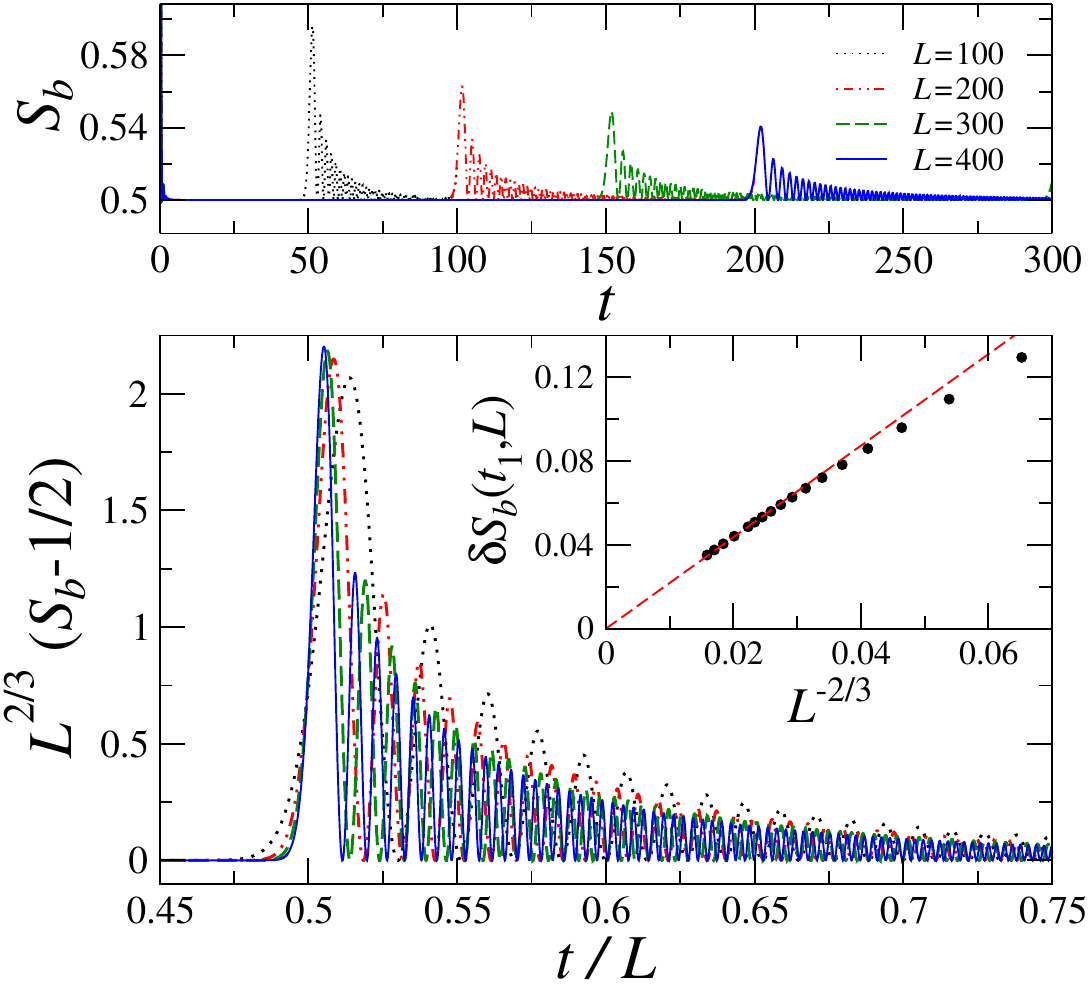}
\caption{Finite-size features of the temporal behavior of the boundary
  transverse magnetization, with OBC, after a quench from
  $g_0=+\infty$ to $g=1$.  The top panel shows curves for $S_b$ versus
  the time $t$, while the bottom panel shows curves for $L^{2/3}
  \delta S_b$ [cf.~Eq.~(\ref{desb})] versus the rescaled time
  $t_L\equiv t/L$.  The inset of the bottom panel highlights the
  scaling behavior of the height of the first peak (located at $t_1
  \approx L/2$), showing data for $\delta S_b(t_1, L)$ vs. $L^{-2/3}$;
  the dashed red line is a linear fit for $L \geq 240$.}
\label{fig:MtOBC_1st_g1}
\end{figure}

The behavior of the boundary transverse magnetization is
substantially different.  Let us define again a subtracted quantity
\begin{equation}
  \delta S_b(t,L) \equiv S_b(t,L) - S_b(t,L\to\infty)
  \label{desb}
\end{equation}
where $S_b(t,L\to\infty)$ can be accurately obtained from the results
for the largest lattices, as one may infer from the curves shown in
the upper panel of Fig.~\ref{fig:MtOBC_1st_g1}.  In particular, we
have verified that the apparent relation $S_b = S_c$ for $t\to\infty$
and $L\to\infty$ actually holds only for $g_0\to\infty$, and, even in
this case, the equality does not hold at finite $t$.  
Finite-size effects are captured by a scaling equation of the form
\begin{equation}
  \delta S_b(t,L) = L^{-b} f_e(t_L) f_o(t,L) + O(L^{-1})\,,  
  \label{asydsb}
\end{equation}
where 
\begin{equation}
 b\approx 2/3\,,
  \label{kappab}
\end{equation}
as emerging from the data reported in the lower panel.  This behavior
is again characterized by the rapid oscillations $f_o(t,L)$ and a
smooth envelope function $f_e(t_L)$ with argument $t_L=t/L$. The main
difference is that such a smooth finite-size behavior gets suppressed
by a power $L^{-2/3}$ (see, in particular, the inset in the lower
panel), and peculiar structures emerge at $t_{L,k} = k /v_m$.

We conclude this part by mentioning that the above behaviors are not
peculiar of the choice of the initial condition, corresponding to the
ground state for $g_0\to\infty$.  Indeed, their main features also
emerge in quenching protocols starting from ground states associated
with finite values $g_0>1$ (data not shown).

\section{Quenches from the disordered phase
in the presence of dissipation}
\label{dissipation}

\subsection{Modeling dissipation}
\label{moddiss}

The purpose of this section is to extend the previous analysis on the
dynamic features of closed quantum Ising chains, in order to include
the effects of weak dissipative mechanisms: besides the
changes of the Hamiltonian parameters, we suppose that the many-body
system is also subject to some interaction with the environment.
For the sake of simplicity, here we only concentrate on quenches
starting from the disordered phase ($|g_0| > 1$), such that it is
still possible to exploit the mapping to the Kitaev Hamiltonian for
fermionic particles~\eqref{kitaev2}, as we did in
Sec.~\ref{fssdis_pbc} and~\ref{fssdis_obc}.  We consider dissipation
mechanisms associated with either particle losses or pumping on each
lattice site, so that our system-bath coupling scheme describes the
coupling of each site with an independent bath.  In the case of weak
coupling to Markovian baths, the general Lindblad master
equation~\eqref{lindblaseq} can be thus written as~\cite{Lindblad-76,
  GKS-76}
\begin{equation}
  {\partial\rho\over \partial t} = -i \big[ \hat H_{\rm K},\rho \big]
  + u \sum_j \Big[ \hat L_j \rho \hat L_j^\dagger - \tfrac{1}{2}
  \big( \rho\, \hat L_j^\dagger \hat L_j + \hat L_j^\dagger 
\hat L_j \rho \big) \Big] \,.
  \label{dLj}
\end{equation}
The onsite Lindblad operators associated to either losses (l) or
pumping (p) are respectively given by~\cite{HC-13, NRV-19, Davies-70,
  Evans-77, SW-10, Nigro-19}:
\begin{equation}
  \hat L^{(\rm l)}_j = \hat c_j \,, \qquad \hat L^{(\rm p)}_j = 
\hat c_j^\dagger \,.
  \label{loppe}
\end{equation}
The choice of such dissipators turns out to be particularly convenient
for the numerical analysis, allowing us to maintain the polynomial
scaling with $L$ of the computational complexity of the problem, as
for the unitary dynamics of the Kitaev chain.

In the rest of this section, for our convenience we shall restrict to
homogeneous dissipation mechanisms and to Kitaev models with ABC
(corresponding to PBC for the Ising chain, when quenching from
ground states in the disordered phase), in such a way to preserve
translational invariance and to further reduce computational resources
to a linear amount in $L$.  We study the time behavior of the
analogue, in fermionic language, of the transverse magnetization with
either incoherent losses or pumping
\begin{equation}
  S^{({\rm l} / {\rm p})}(t,g_0,g,u,L) =
    2 \, {\rm Tr} \big[ \hat c^\dagger_j \hat c_j \, \rho(t) \big] -1 \,,
\end{equation}
where the superscripts $({\rm l})$ and $({\rm p})$ refer to the loss
or pumping dissipations related to the Lindblad operators
(\ref{loppe}).  Note that the initial condition $|\Phi_{g_0}\rangle$,
being the ground state of $\hat H_{\rm K}$, can be interpreted as a
state with a local fermionic filling given by $n_j = \langle
\Phi_{g_0}| \hat c^\dagger_j \hat c_j |\Phi_{g_0}\rangle$, with $n_j
\in [0,1]$.  For example, the extreme cases of disordered ground
states are denoted by the completely filled state of fermions
$|\Phi_{+\infty}\rangle = |1, \ldots, 1\rangle$ and the completely
empty state of fermions $|\Phi_{-\infty}\rangle = |0, \ldots,
0\rangle$.

\begin{figure}%[!t]
  \includegraphics[width=0.95\columnwidth]{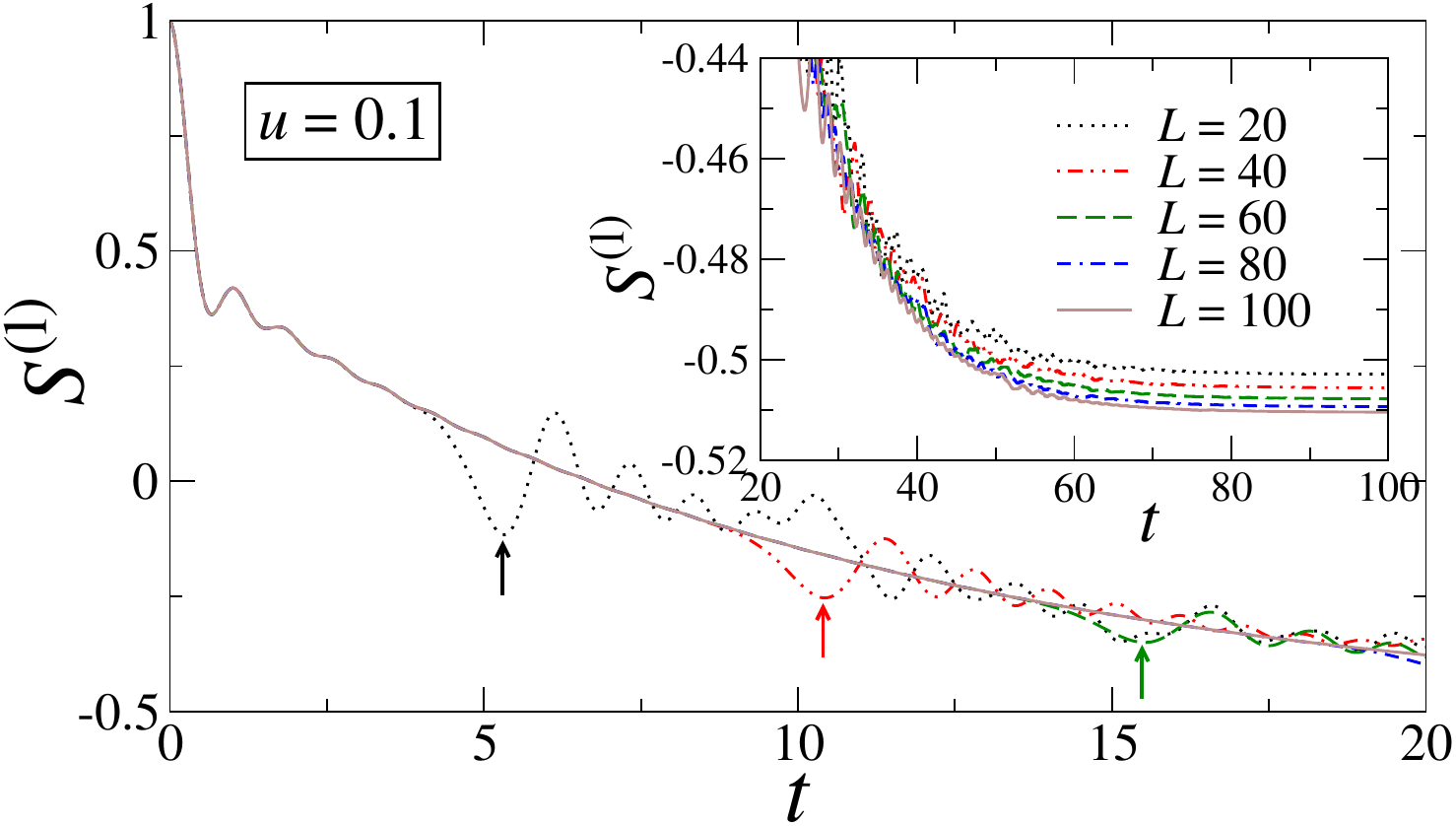}
  \includegraphics[width=0.95\columnwidth]{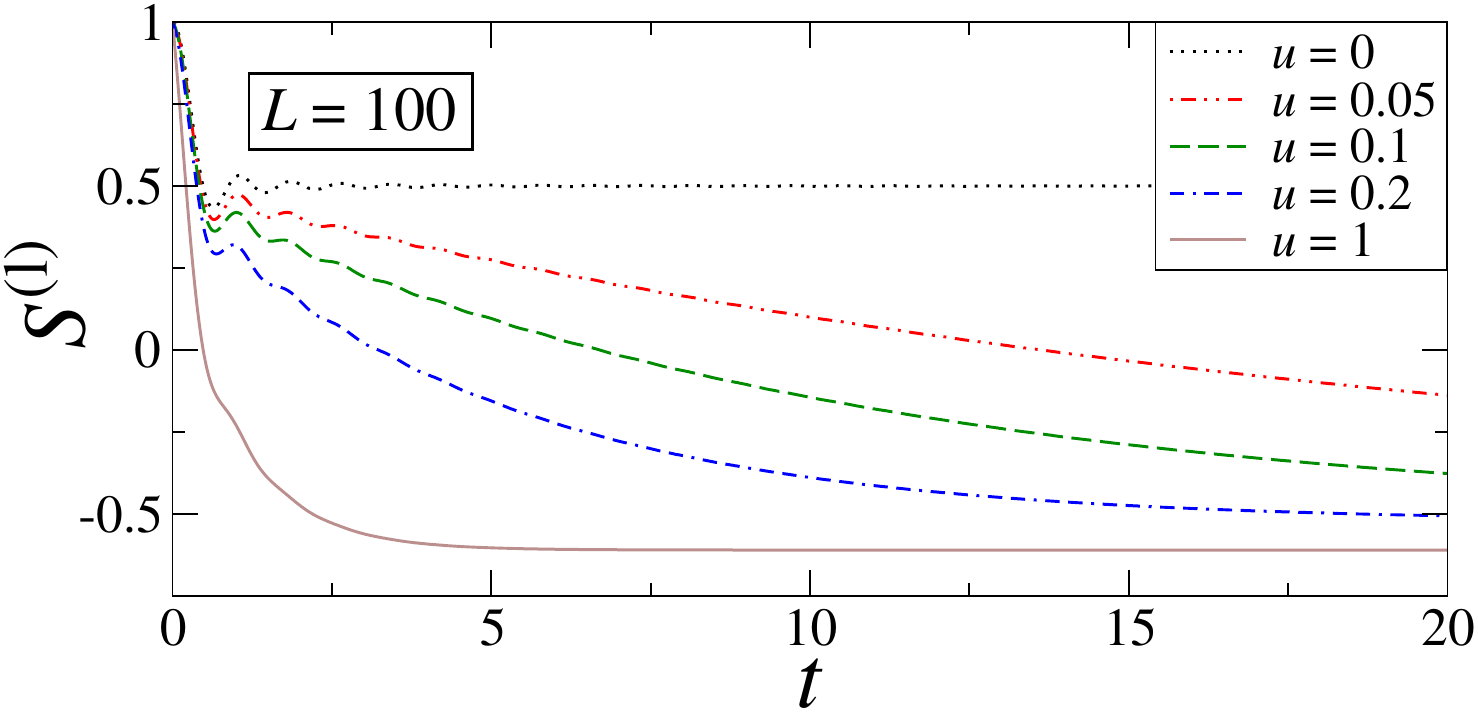}
  \caption{Time behavior of the transverse magnetization $S^{(\rm l)}$
    after a quench in the Kitaev quantum wire, starting from a
    completely filled state ($g_0 = + \infty$), to $g=g_c=1$ (critical
    point), in the presence of incoherent particle losses.  Upper
    panel: curves are for different sizes $L$, at fixed dissipative
    coupling constant $u=0.1$ (the inset shows a magnification at long
    times); arrows indicate the position $t_1$ of the first dip that
    is ascribable to finite-size effects. Lower panel: curves
    correspond to different values of $u$ and fixed $L=100$.}
\label{fig:dissipation}
\end{figure}

In Fig.~\ref{fig:dissipation} we show some representative examples of
the time dependence of the transverse magnetization when the system
evolves according to the Lindblad equation~\eqref{lindblaseq} for
$g=1$.  We start from the completely filled state corresponding to the
ground state for $g_0\to+\infty$, so that $S(t=0)=1$. The dissipation
is modeled in the form of particle losses, whose action is to deplete
the system, and thus the value of $S$ tends to generally decrease in
time.  The upper panel displays the time behavior of $S^{(\rm l)}$ for
different wire lengths $L$ and fixed dissipation strength $u = 0.1$.
With increasing $L$, the various curves overlap up to progressively
longer times: for $L \geq 20$ all of them appear to be identical up to
$t \lesssim 3$, while looking at the curves for larger sizes $L=80$
and $100$, we see that they cannot be distinguished up to much longer
times ($t \approx 18$).  This fact witnesses the existence of a well
defined temporal behavior in the thermodynamic limit $L\to \infty$, as
already pointed out for the unitary dynamics of a closed system (see
Fig.~\ref{fig:Kitaev_t_g1}).  In fact, the emergence of dips and
wiggles at finite size $L$, departing from the large-$L$ behavior and
whose position progressively shifts to later times with increasing $L$
(see arrows highlighting the first of such dips, clearly visible at
the smallest available sizes), can be ascribed to finite-size effects,
as we will see in Sec.~\ref{dissfss}.  Note also the emergence of an
asymptotic stationary value $S^{(\rm l)}(t=\infty) \approx -0.5$,
which weakly depends on the value of $L$, as hinted by the data
reported in the inset.
The bottom panel of Fig.~\ref{fig:dissipation} shows how the
progressive increase of the dissipation strength $u$ leads to a faster
decay of $S^{({\rm l})}$ in time, down to the asymptotic behavior.
This contrasts with the unitary case ($u=0$), where $S$ converges to
the large-time limit according to the oscillating behavior predicted
by Eq.~\eqref{ftgg0inf}.

Below we focus on all these issues in more detail, specifically
addressing the thermodynamic limit behavior and finite-size
corrections.  As we shall see in the following analysis, some features
characterizing the unitary evolution of the closed system disappear,
while other will leave some residual trace.

\begin{figure*}[!t]
  \includegraphics[width=0.95\textwidth]{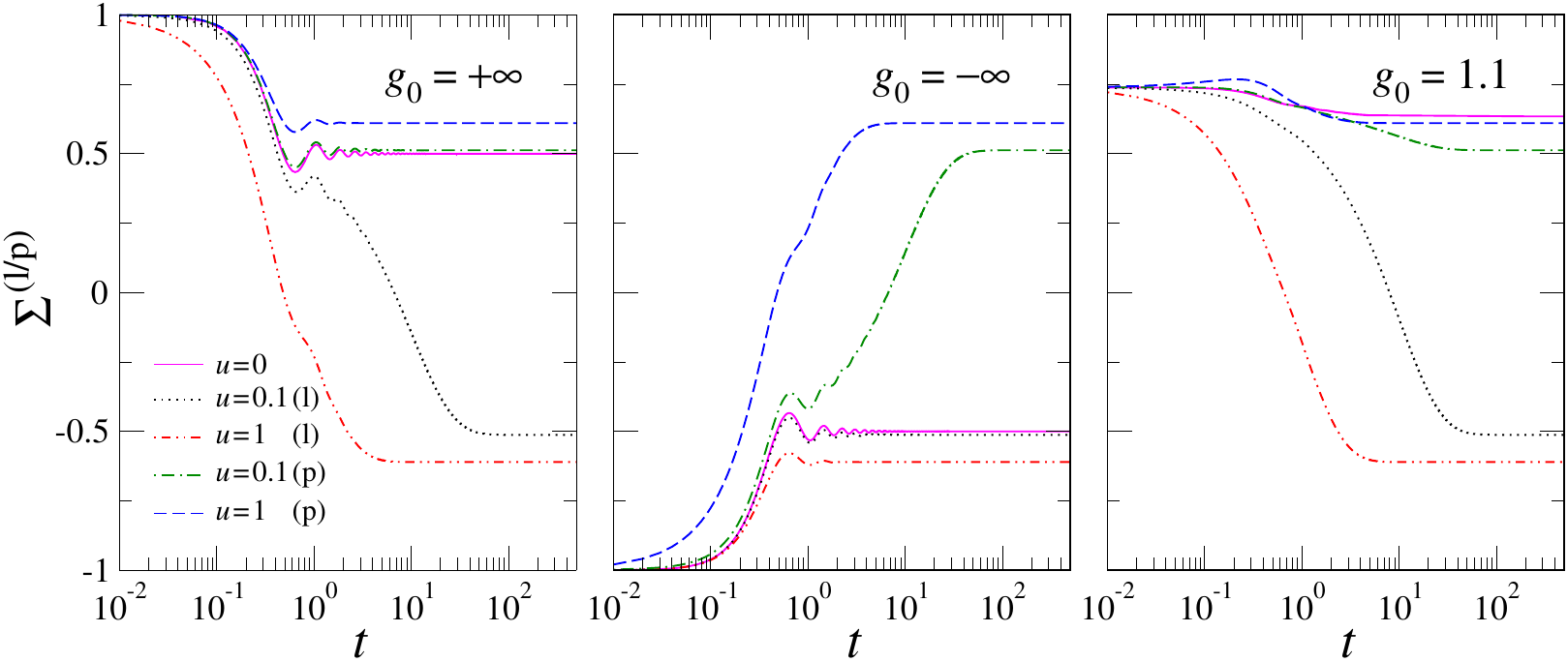}
  \caption{Transverse magnetization as a function of time, in the
    dissipative Kitaev quantum wire quenched to the critical point
    $g=1$.  The three panels refer to different initial conditions,
    corresponding to: a completely filled state ($g_0 = +\infty$,
    left), a completely empty state ($g_0=-\infty$, central), and a
    partially filled state ($g_0=1.1$, right).  The dissipation is
    implemented in the form of either particle losses (l) or pumping
    (p), with two different strengths (see legend).  The continuous
    purple line in each panel denotes the behavior in the absence of
    dissipation ($u=0$), cf.~Eq.~\eqref{asymt}, which converges to
    $F(g,g_0)$ for $t\to\infty$ [see Eq.~\eqref{fagg0}].  In all the
    curves reported here we have fixed $L=1000$, after checking that
    such system size was sufficiently long to guarantee the study of
    time dependence in the thermodynamic limit. Times are in
    logarithmic scale, to better highlight the convergence to the $t
    \to \infty$ stationary behavior.}
\label{fig:Mt_KitaevDiss_t}
\end{figure*}

\subsection{Behavior in the thermodynamic limit}
\label{dissthlimit}

Here we consider sufficiently large system sizes, so that finite-size
effects are guaranteed not to play any role (i.e., for the times
analyzed below, curves obtained for the two largest available values
of $L$ do overlap at any time).  Note however that, due to the
presence of dissipation, revivals are generally expected to be
suppressed, and the influence of the boundary conditions to be less
important than in closed systems.  In practice, we carefully checked
that chains of length $L=1000$ were sufficiently long to ensure that
we are always probing the thermodynamic limit behavior (for example
by comparing results for different sizes, and requiring agreement).

Analogously to closed systems, we define the thermodynamic limit of
the transverse magnetization
\begin{equation}
\Sigma^{({\rm l/p})}(t,g_0,g,u) \equiv S^{({\rm l/p})}(t,g_0,g,u,L\to\infty) 
\,, \label{asymtu}
\end{equation}

When monitoring the time evolution of the transverse magnetization
after a quench of the Hamiltonian parameter $g$ and switching on the
dissipation at $t>0$, one finds a behavior of the type reported in
Fig.~\ref{fig:Mt_KitaevDiss_t}: $\Sigma^{(\rm l/p)}$ starts from its
corresponding value at the ground state with $g_0$, and then evolves
in time until it converges to an asymptotic steady-state value, for $t
\to \infty$.  Comparing the three panels for different initial
conditions, being either the completely filled, completely empty, or a
partially filled state, we note that the steady-state magnetization
$\Sigma^{(\rm l/p)}(t\to\infty)$ does not depend on such choice.  This
is a particular feature related to the presence of dissipation.
Indeed the uniqueness of the (possibly existing) steady state has been
proven for the loss and pumping operators~\cite{Davies-70, Evans-77,
  SW-10, Nigro-19}.  This contrasts what happens in the absence of
dissipation, where the surviving time-independent contribution
$F(g_0,g)$ clearly depends both on $g$ and on $g_0$ [see
  Eq.~\eqref{fagg0} and the plateau reached at long times by the
  continuous purple lines in Fig.~\ref{fig:Mt_KitaevDiss_t}].

Notice however that, in the presence of incoherent pumping, the
transverse magnetization evolves in time similarly to a closed system,
which is initialized in the completely filled state
$|\Phi_{+\infty}\rangle$ (left panel). Conversely, with incoherent
decay, $\Sigma$ behaves similarly to a closed system initialized in
the completely empty state $|\Phi_{-\infty}\rangle$ (central panel).
Asymptotic long-time values are also relatively close to the
non-dissipative predictions given by $F(\pm \infty,1)$, respectively.
While the specific time dependence of $\Sigma^{({\rm l/p})}$ depends
on the dissipation strength $u$, discrepancies with respect to the
corresponding unitary behavior tend to amplify with increasing $u$
(compare the continuous curves at $u=0.1$ with those at $u=1$), and
even at $u=1$ it is possible to recognize a qualitatively similar
trend.
The above similarities between unitary and dissipative dynamics
disappear when starting from a partially filled initial state
$|\Phi_{g_0, \, {\rm with} \, |g_0|<\infty}\rangle$ (right panel).

As emerging from the data presented in Fig.~\ref{fig:Mt_KitaevDiss_t},
the time evolution of the transverse magnetization satisfies
\begin{equation}
  \Sigma^{(\rm l)}(t,\pm \infty,g,u) = -\Sigma^{(\rm p)}(t,\mp
  \infty,g,u) \,,
\end{equation}
when starting from one of the extremal states $|\Phi_{\pm \infty}
\rangle$ (compare left and central panels).  The same holds for the
unitary dynamics obtainable setting $u=0$, i.e.
$\Sigma(t,+\infty,g,0) = - \Sigma(t,-\infty,g,0)$.  This is no longer
the case if $|g_0|<\infty$. However, due to the independence on $g_0$
of the steady-state value of $\Sigma$, we still have for fixed
dissipation strength $u$:
\begin{equation}
  \Sigma^{(\rm l)}(t\to\infty,g_{0{\rm l}},g,u) = 
-\Sigma^{(\rm p)}(t\to\infty,g_{0{\rm p}},g,u),
  \label{symm_mt_asynt}
\end{equation}
independently of the values $g_{0{\rm l}}$ and $g_{0{\rm p}}$.

We stress that, while the figure only reports results for quenches
ending at the critical point $g=g_c=1$, analogous conclusions on the
steady-state values and symmetry properties of the model can be drawn
for generic values of $g$ (not shown here).

\begin{figure}%[!t]
  \includegraphics[width=0.95\columnwidth]{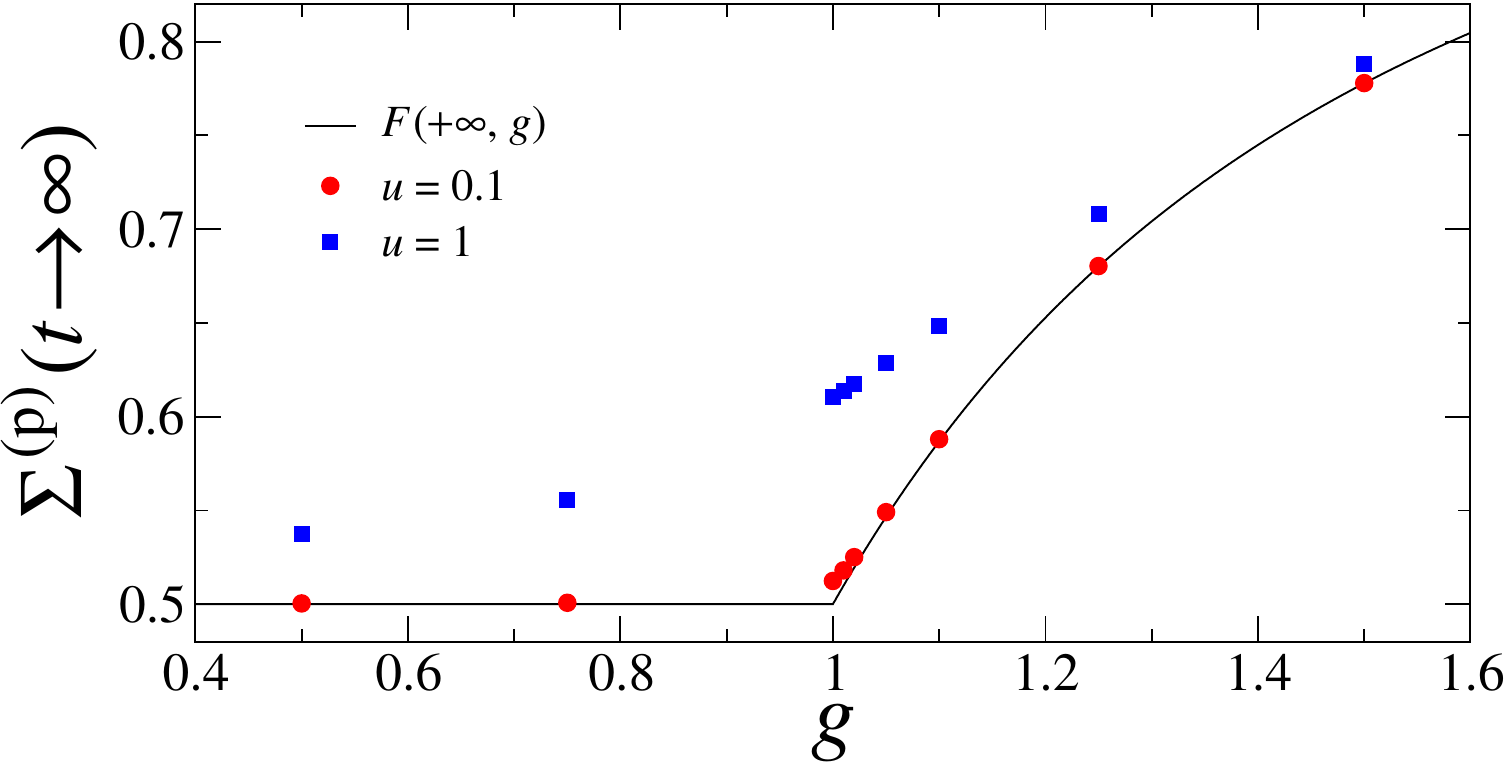}
  \caption{Asymptotic long-time value reached by the transverse
    magnetization, $\Sigma^{\rm(p)}(t\to\infty)$, as a function of the
    post-quench value of $g$ in the presence of incoherent pumping of
    strength $u=0.1$ (red circles) and $u=1$ (blue squares).  The
    continuous black line is the function $F(+\infty,g)$ of
    Eq.~\eqref{fagg0inf} for the large-time limit in the absence of
    dissipation and starting from $g_0 = +\infty$.}
\label{fig:Mt_KitaevDiss_asynt}
\end{figure}

To better analyze the influence of dissipation and of the post-quench
transverse field $g$ on the long-time value of the transverse
magnetization, in Fig.~\ref{fig:Mt_KitaevDiss_asynt} we report the
dependence of $\Sigma^{(\rm p)}(t\to\infty)$ on $g$, for two distinct
values of incoherent pumping strength $u$.  These are compared with
the unitary prediction given by $F(+\infty,g)$
[cf.~Eq.~\eqref{fagg0inf}].  As is visible, the discontinuity in the
first derivative at $g_c$ is smeared by the coupling with the
environment.  For small $u$ it is however possible to recover a signal
of singularity in $g_c$.  In the figure we did not report the case of
incoherent decay, being equal to the opposite value of pumping [see
  Eq.~\eqref{symm_mt_asynt}] and matching the corresponding unitary
prediction $F(-\infty,g)$.

\begin{figure}%[!t]
  \includegraphics[width=0.95\columnwidth]{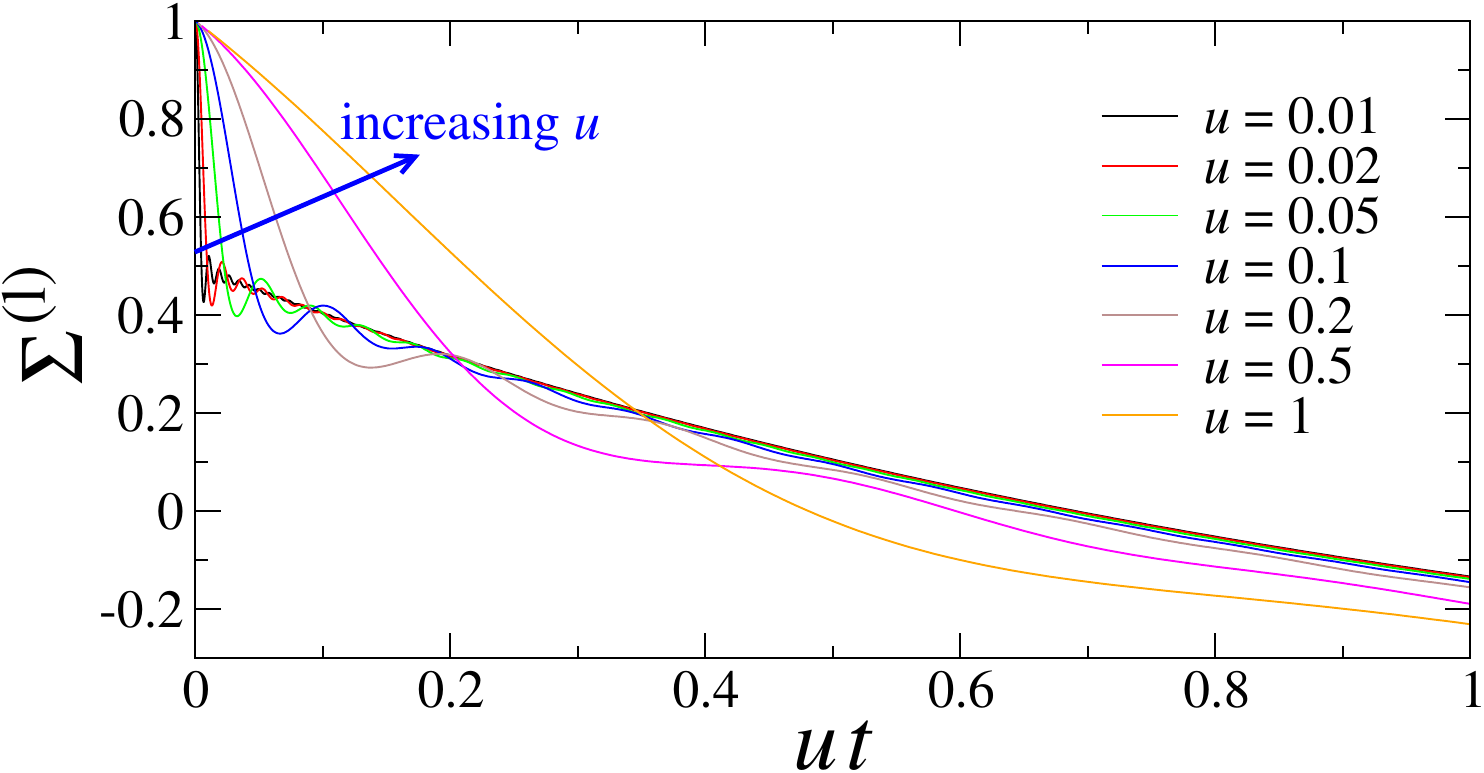}
  \caption{Scaling behavior of $\Sigma^{\rm (l)}$ in the presence of
    incoherent particle losses, for a quench from $g_0=+\infty$ to
    $g=1$.  The various curves are for different values of $u$ (see
    legend) and $L=1000$, so to ensure having reached the behavior in
    the thermodynamic limit. Times have been rescaled according to the
    adimensional variable $\nu = u \, t$.}
\label{fig:Mt_KitaevDiss_rescal_t}
\end{figure}

We conclude this part by observing that, in the large time-limit, the
time evolution of the transverse magnetization, for small dissipation
strength $u \ll 1$, obeys an asymptotic scaling behavior of the form
\begin{equation}
  \Sigma^{(\rm l/p)} (t, g_0 , g, u) 
\approx \Sigma^{(\rm l/p)}_r(g_0,g, \nu)\,,
  \qquad \nu \equiv u\, t\,,
  \label{eq:Sigmasca}
 \end{equation}
keeping $\nu$ fixed.  The data in
Fig.~\ref{fig:Mt_KitaevDiss_rescal_t}, for various values of the
strength $u$ of incoherent losses, nicely support this behavior (note
also that, for $u \gtrsim 0.1$, the collapse of the various curves
definitely becomes less accurate). An analogous data collapse is seen
in the case of incoherent particle pumping.  We stress that the
scaling behavior put forward in Eq.~\eqref{eq:Sigmasca} should not be
considered as unexpected, since the parameter $u$ plays the role of
decay rate of the dissipation.

\subsection{Finite-size effects}
\label{dissfss}

\begin{figure}%[!t]
  \includegraphics[width=0.95\columnwidth]{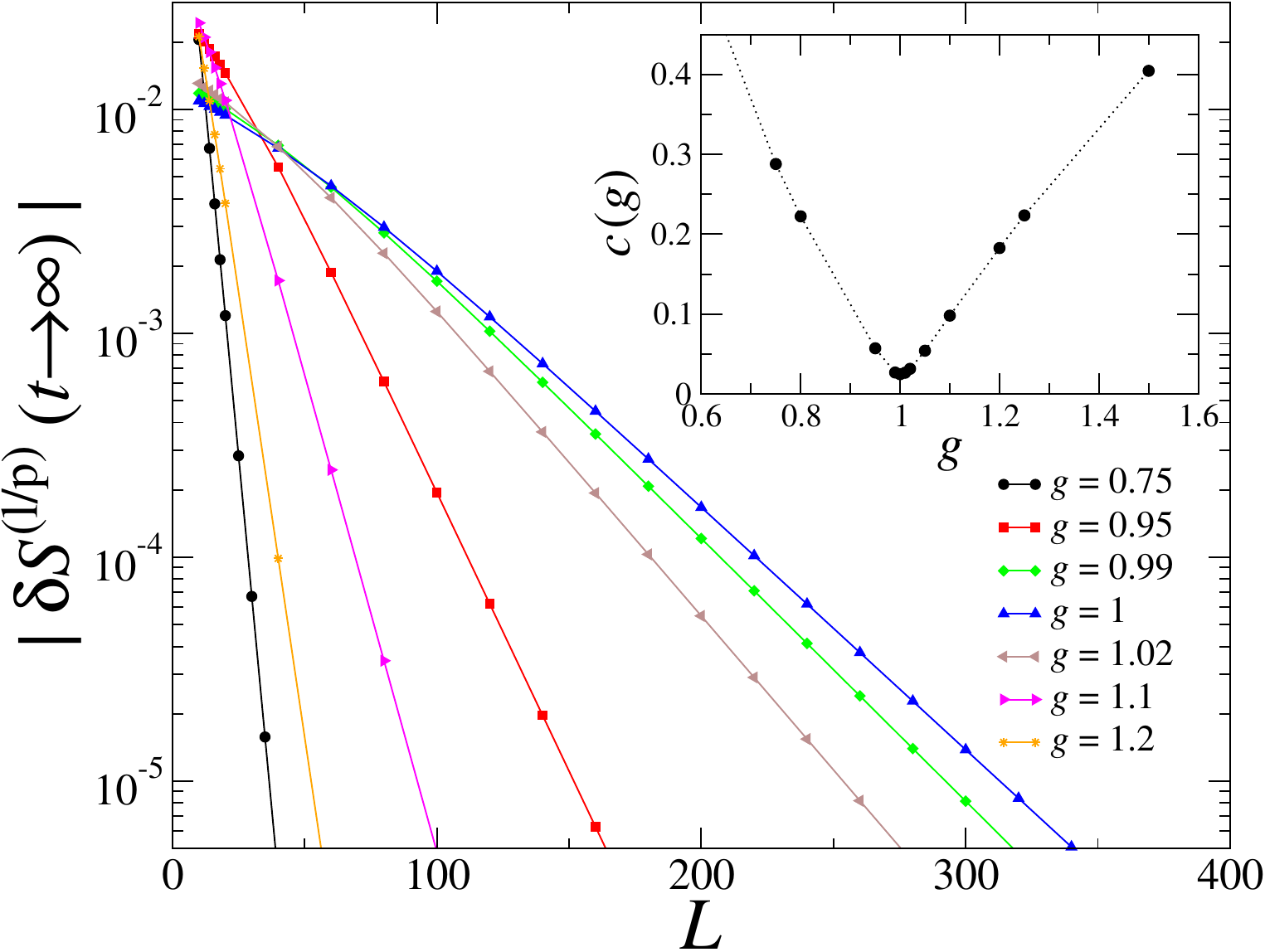}
  \caption{The difference $\delta S^{\rm (l/p)}$,
    cf. Eq.~\eqref{deltaslp}, in the asymptotic long-time limit,
    between the transverse magnetization at finite length $L$ and the
    corresponding thermodynamic limit value for $L\to \infty$, The
    various curves are for different values of $g$ and for fixed
    dissipation strength $u=0.1$.  These data are unaffected by the
    choice of dissipation (being either losses or pumping) and of
    $g_0$.  The inset shows the exponential decay rate of such
    discrepancy, as a function of $g$.}
\label{fig:Mt_KitaevDiss_L}
\end{figure}

As for the case of closed systems, we now focus on finite-size effects
for PBC and look at the difference
\begin{equation}
  \delta S^{(\rm l/p)}\!(t,g_0,g,u,\!L) \! = \!
  S^{(\rm l/p)}\!(t,g_0,g,u,\!L) - \Sigma^{\rm (l/p)}\!(t,g_0,g,u)\,,
  \label{deltaslp}
\end{equation}
where $\Sigma^{\rm (l/p)}\!(t,g_0,g,u)$ is the infinite-size limit of
$S^{(\rm l/p)}\!(t,g_0,g,u,\!L)$.
Figure~\ref{fig:Mt_KitaevDiss_L} displays such difference in the
asymptotic long-time limit for a prototypical dissipative situation,
where the system is quenched to the critical point $g_c$ and with
dissipation strength $u=0.1$.  As noted before, the value reached by
$\delta S$ in the long-time stationary regime does not depend on $g_0$
and on the type of dissipation (losses or pumping).  The various
curves show that $\delta S^{(\rm l/p)}$ decays exponentially in the
large-$L$ limit, i.e.
\begin{equation}
  \delta S^{(\rm l/p)}(t\to\infty) \sim e^{- c(g) L}\,.
  \label{expbeh}
\end{equation}
It is also interesting to note that the exponential rate $c(g)$
displays a minimum at $g=g_c=1$, thus signaling the emergence of
larger finite-size corrections at the critical point (although they
are still exponentially suppressed).

\begin{figure}%[!t]
  \includegraphics[width=0.95\columnwidth]{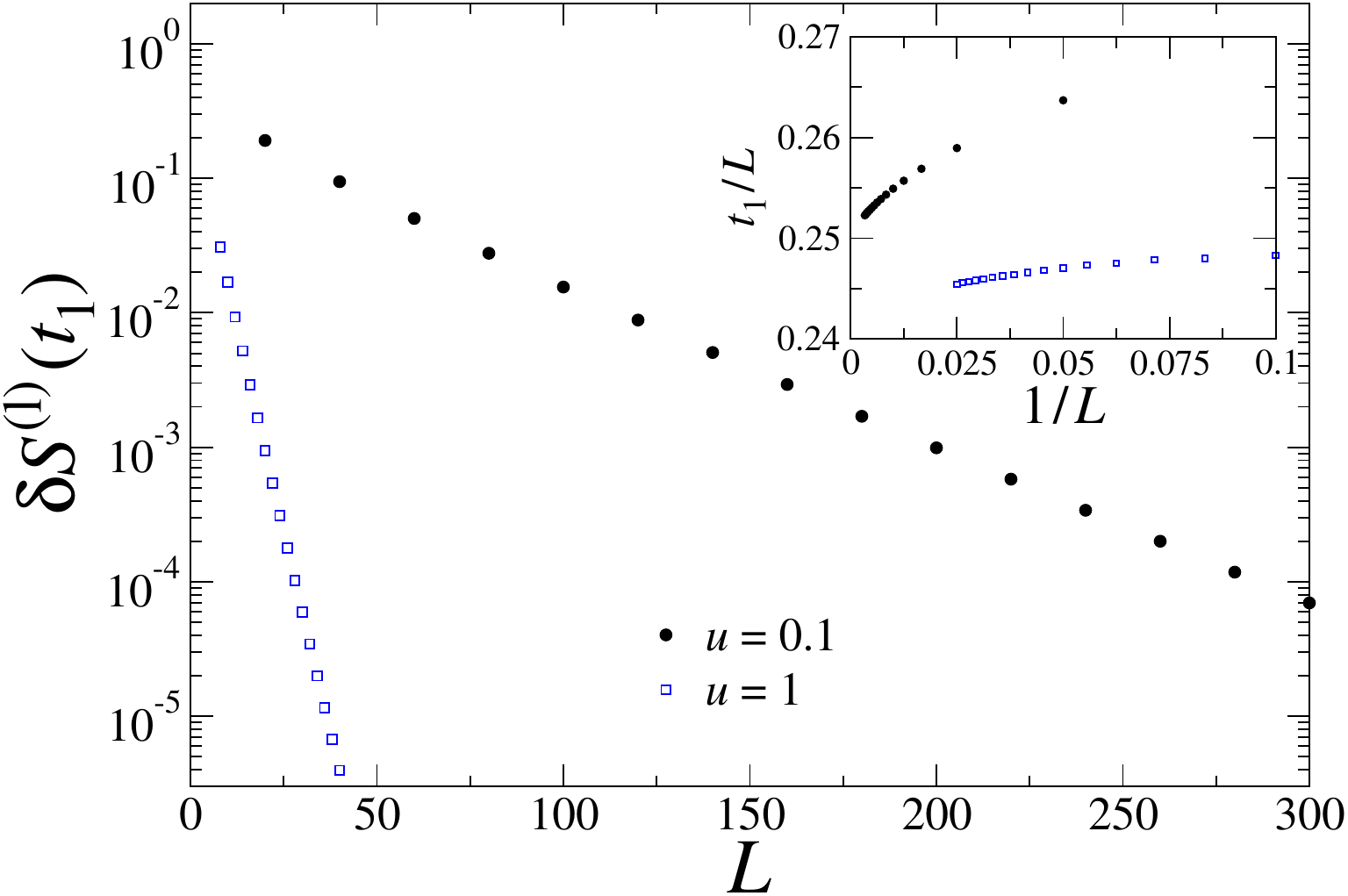}
  \caption{Scaling behavior of the first dip (see arrows in
    Fig.~\ref{fig:dissipation}) corresponding to finite-size effects
    in the transverse magnetization, for a quench from $g_0=+\infty$
    to $g=1$ and in the presence of incoherent losses of strength
    $u=0.1$.  The main frame displays the value of the subtracted
    magnetization at the dip, $\delta S^{\rm (l)}(t_1)$
    [cf.~Eq.~\eqref{deltaslp}], as a function of $L$ in semilog
    scale. The inset displays the position of the dip $t_1/L$
    vs.~$1/L$.}
\label{fig:dissipation2}
\end{figure}

Coming back to the time behavior of $S^{\rm (l)}(t)$ at finite size,
shown in the upper panel of Fig.~\ref{fig:dissipation}, we can observe
a dip (and subsequent wiggles) at larger and larger times $t_1$ with
increasing $L$, so that $t_1\sim L$ (see the inset of
Fig.~\ref{fig:dissipation2}).  This finite-size behavior apparently
resembles the one already observed in closed systems, as in
Fig.~\ref{fig:Kitaev_t_g1}.  However, a closer inspection of the
scaling of the height of such dips reveals that their depth is
exponentially suppressed with increasing $L$, as clearly visible from
the data shown in the main frame of Fig.~\ref{fig:dissipation2}.  The
large-$L$ limit is thus approached much faster than for the unitary
dynamics (compare, e.g., with Fig.~\ref{fig:Peak_XY_Kitaev_g1}).  We
checked that these facts are not qualitatively affected by the choice
of the pre- and post-quench values $g_0$ and $g$, nor by the type and
the strength of dissipation.

In conclusion we observe that, differently from the unitary dynamics
of closed systems, the presence of dissipation dramatically reduces
finite-size effects and no relevant revival phenomena emerge, as
substantially expected.

\section{Quenches from the ordered state}
\label{quordsta}

We now address quenches of the quantum Ising chain~\eqref{hedef}
starting from a completely ordered state (i.e., a fully magnetized
state $|\Psi(t=0)\rangle = |\!\! \uparrow, \ldots , \uparrow
\rangle$).  This corresponds to one of the degenerate ground state in
thermodynamic limit when $g_0\to 0$, obtainable by the ground state in
the limit
\begin{equation}
  |\Psi(0)\rangle = \lim_{g_0\to 0} \lim_{h\to 0^+} \lim_{L\to\infty} | 
  {\rm GS} \rangle \,,
  \label{gsdeg}
\end{equation}
where $h$ is an external homogenous magnetic field coupled to the
longitudinal magnetization.  Since such initial state breaks the
$\mathbb{Z}_2$ symmetry~\eqref{z1sym} of the model, one finds a
nonzero longitudinal magnetization $M$ [cf.~Eqs.~\eqref{mlj} and
  \eqref{mndef}] along the quantum evolution~\eqref{unitary} after
quenching the transverse field parameter $g$.  We will not explicitly
show data for initial partially ordered states (i.e., corresponding to
finite values of $0<g_0<1$), since the same conclusions apply also in
such circumstances. Therefore, in the following we drop the dependence
on $g_0$ of the observables.

\begin{figure}[!t]
  \includegraphics[width=0.95\columnwidth]{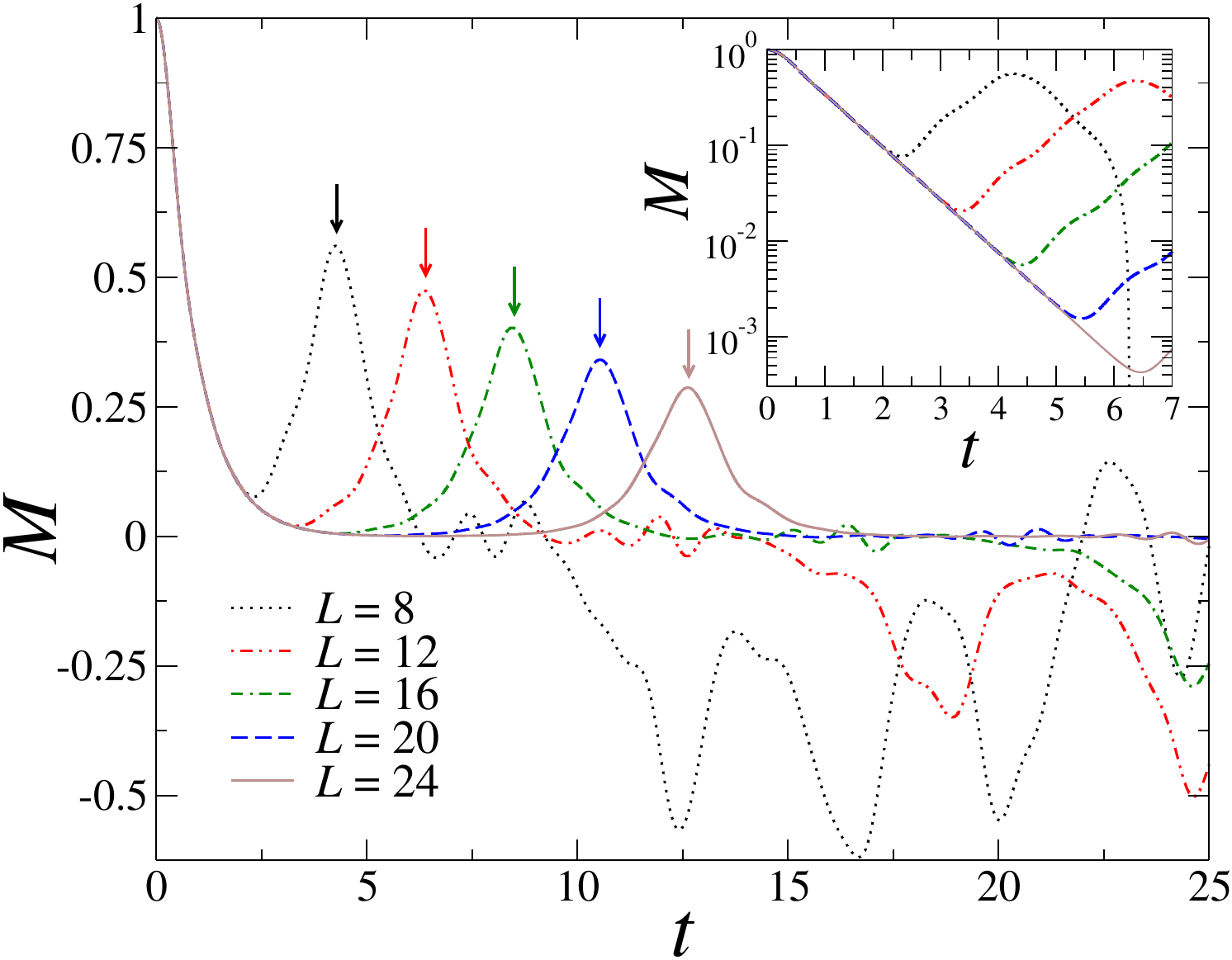}
  \caption{Longitudinal magnetization after a quench in the quantum
    Ising chain with PBC, starting from the symmetry-broken ordered
    state at $g_0 = 0$, to $g=1$.  The various data sets are for
    different chain lengths; arrows denote the position of the first
    peak emerging in the finite-size revivals, located at time
    $t_1(L)\approx L/(2v_m)$.  The inset is a magnification at small
    times in semilog scale, showing that, in the thermodynamic limit,
    the magnetization decays exponentially with $t$.}
\label{fig:Ml_t_g1}
\end{figure}

An example of the evolution of $M(t)$ after the above protocol is
provided by Fig.~\ref{fig:Ml_t_g1}, where we focus on quenches to the
critical point $g=g_c=1$ for different lattice sizes $L$.  Similarly
to the case of quenches from disordered states, the small-time
behavior appears to be independent of the size (see, e.g.,
Fig.~\ref{fig:Kitaev_t_g1}), and in this case it drops exponentially
in time.  Then, after some time which depends on $L$ and on the choice
of the boundary conditions, a non monotonic behavior related to
finite-size revival effects sets in, as we shall clarify below.

At this stage it is important to stress that, for quenches from
ordered states ($g_0 < 1$), one cannot exploit the exact mapping with
the quantum fermionic wires~\eqref{kitaev2} with ABC. Therefore in
this case, to study the exact out-of-equilibrium behavior of the
system, we are forced to employ exact-diagonalization methods which
typically scale exponentially with $L$ and thus severely limit the
simulatable lattice sizes up to $L \simeq 24$.  In passing we mention
that a new formalism based on the correspondence between momentum
space and real space has been recently put forward in
Ref.~\cite{Wu-20}, which would enable to evaluate the longitudinal
magnetization for considerably larger system sizes, by numerically
computing suitable Pfaffians.  Nonetheless, as we shall see in a
moment, brute-force diagonalization will be sufficient to infer the
main dynamical features we are interested in.

\subsection{Quenches in the thermodynamic limit}
\label{qudisthlim}

We first discuss the thermodynamic limit of the longitudinal
magnetization $M(t)$.  For the transverse magnetization $S(t)$, it is
sufficient to take the limit $g_0\to 0$ of the exact analytic formulas
reported in Sec.~\ref{thtrma}.

The longitudinal magnetization in the thermodynamic limit, 
\begin{equation}
  {\cal M}(t,g) \equiv M(t,g,L\to\infty) \,,
  \label{mlthlim}
\end{equation}
turns out to vanish in the large-time limit $t\to\infty$.  Indeed it
presents an asymptotic exponential decay, according
to~\cite{CEF-12-1}
\begin{equation}
  {\cal M}(t,g) \approx {\cal M}_{a}(t,g) = A(g)\, \exp[-\Gamma(g)\,t ]\,,
  \label{mlgg0}
\end{equation}
where
\begin{eqnarray}
  A(g) & = & {1\over \sqrt{2}} \left[ 1 + \sqrt{1-g^2}\right]^{1/2}\,,
  \label{cg}\\
  \Gamma(g) & = & \int_0^\pi {dk\over \pi} \, 2 \:{g \sin(k) \over
    \Lambda(g,k)} \, \ln \left[ {1 - g \cos(k)\over
      \Lambda(g,k)}\right]\,,\quad
  \label{gagg0}
\end{eqnarray}
where $\Lambda(g,k)$ is the function reported in Eq.~(\ref{lagk})
setting $\gamma=1$. The asymptotic behavior (\ref{mlgg0}) was derived
essentially for $g<1$ in Ref.~\cite{CEF-12-1}.
Similarly as for the transverse magnetization [see Eq.~\eqref{fag1}],
even in this case we note a singular behavior at $g=g_c=1$. Indeed,
around $g_c$, the decay function $\Gamma(g)$ behaves as
\begin{equation}
  \Gamma(g) = \left\{ \begin{array}{ll}
    4/\pi - 2\sqrt{2(1 - g)} + O(1-g) & {\rm for} \; g<1 \,,  
\vspace*{2mm} \\
    4/\pi & {\rm for} \; g\ge 1 \,. 
  \end{array} \right. \label{gammag1beh}
\end{equation}
We have carefully checked that these asymptotic analytic behaviors are
supported by our data (not shown).

\begin{figure}[!t]
  \includegraphics[width=0.95\columnwidth]{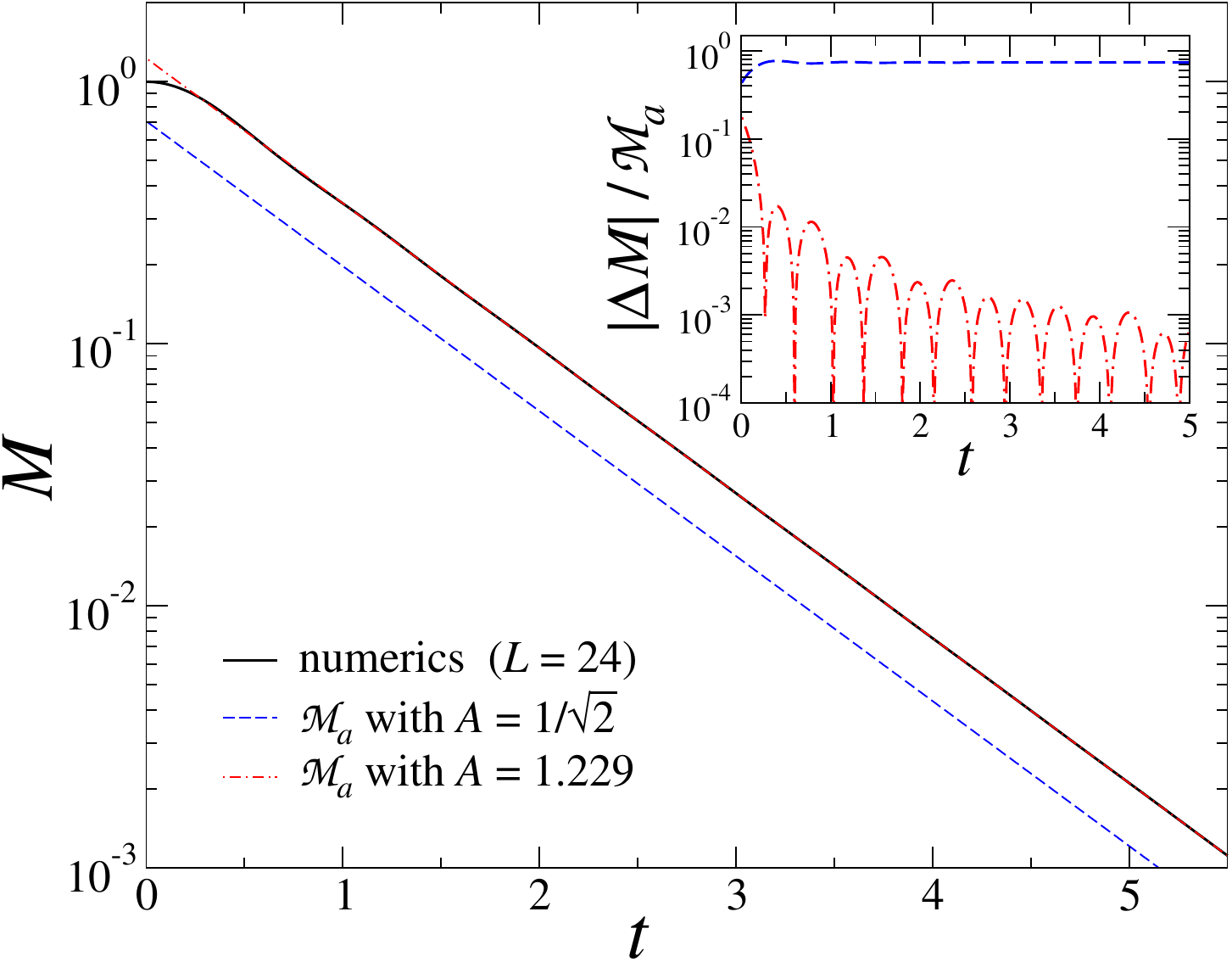}
  \caption{Longitudinal magnetization $M(t)$ in the Ising chain for a
    quench from the completely ordered state to $g=1$. These
    analytical and numerical results are supposed to show the behavior
    in the thermodynamic limit.  In particular, the continuous black
    curve is the result of a numerical simulation with $L = 24$ sites;
    we checked that such system size guarantees the study of time
    dependence in the thermodynamic limit, up to time $t=5.5$.  Dashed
    blue and dot-dashed red data sets respectively denote the
    predictions ${\cal M}(t,1)$ given by Eq.~\eqref{mlgg0} with either
    $A(1)$ in Eq.~\eqref{cg} or with $A(1) = 1.229$.  The inset shows
    the relative difference $|\Delta {\cal M}(t,1)|/{\cal M}_a(t,1)$,
    where $\Delta {\cal M}(t,g) = {\cal M}(t,g) - {\cal M}_a(t,g)$ as
    a function of time, for the two predictions discussed before.}
  \label{fig:Check_analyt_Ml}
\end{figure}

A noteworthy issue is the fact that numerical results evidence a
discontinuity at $g=g_c=1$ also in the prefactor $A(g)$.  As a matter
of fact, while for $g \neq 1$ the formula~\eqref{cg} appears to be
correct, the value $A(1)$ which captures the long-time behavior of $M$
is not provided by the $g \to 1$ limit of Eq~\eqref{cg}.  Indeed, as
explicitly reported in Fig.~\ref{fig:Check_analyt_Ml}, we found that
\begin{equation}
  A(1) \approx  1.229
  \label{a1value}
\end{equation}
(with an accuracy of about two per mille), definitely different from
the prefactor predicted by Eq.~\eqref{cg},
i.e. $A(1)=1/\sqrt{2}\approx 0.707$.  This signals a discontinuity in
the prefactor of Eq.~(\ref{mlgg0}) of the asymptotic large-time
behavior, indicating that such an asymptotic behavior is not uniformly
approached for $g\to 1^-$.  In particular, the analytic estimate with
$A(1)$ given by Eq.~\eqref{a1value} is nearly indistinguishable from
the numerical data, thus demonstrating convergence for times
sufficiently small to ensure the reaching of the thermodynamic limit
for the largest available size (compare the continuous black line with
the dot-dashed red line).  A closer look at the relative
discrepancies shows that, in fact, the exact solution presents an
oscillating contribution that vanishes at long times (see inset).  On
the other hand, the estimate provided by Eq.~\eqref{cg} does not
provide the correct result (dashed blue line).

\begin{figure}%[!t]
  \includegraphics[width=0.95\columnwidth]{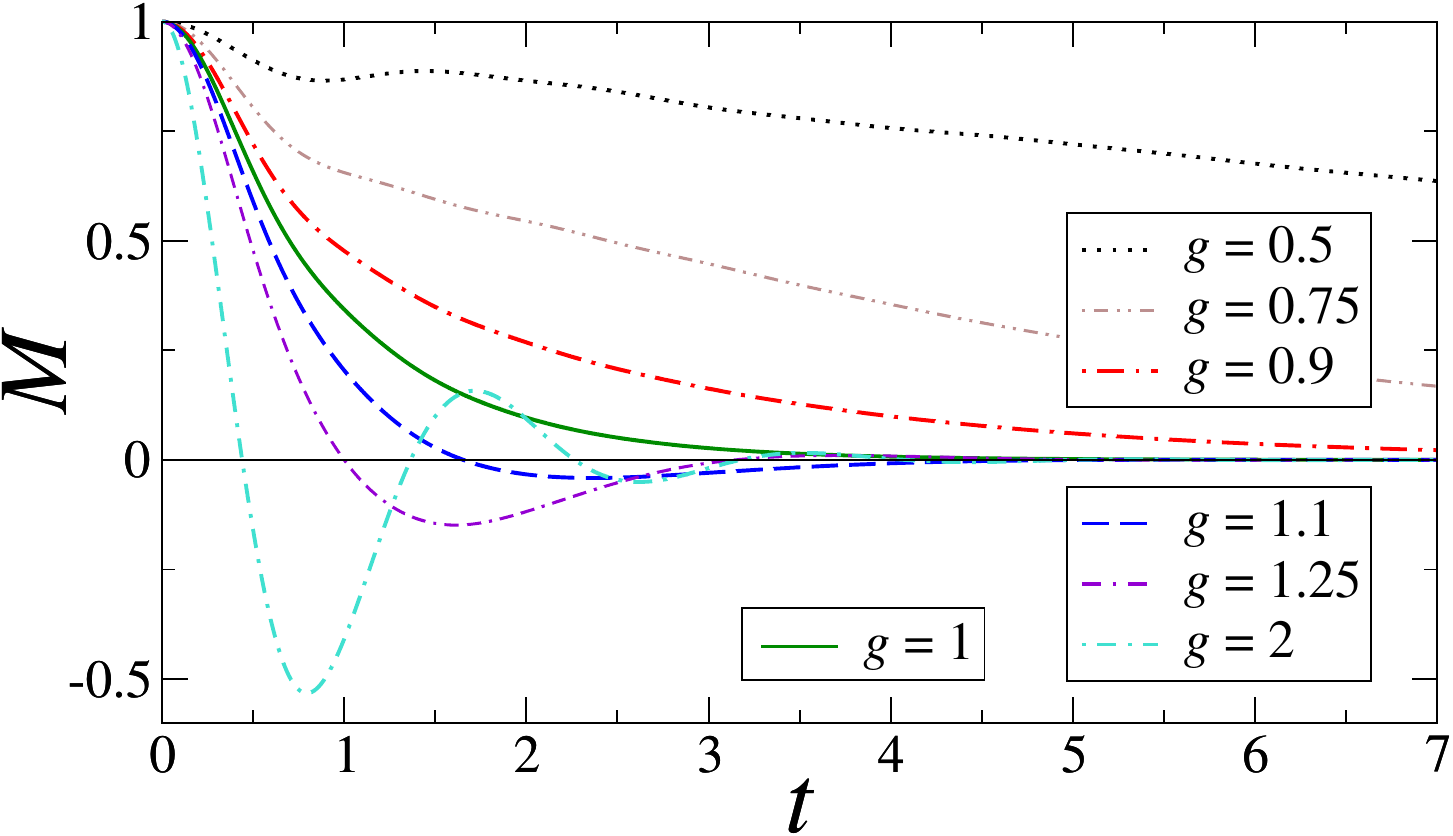}
  \caption{Time dependence of the longitudinal magnetization in the
    thermodynamic limit, up to time $t=7$, after quenches
    to various values of $g$, starting from the fully ordered state.}
  \label{fig:Ising_l_gvar_asynt}
\end{figure}

Let us now have a closer look at the infinite-size limit of the
longitudinal magnetization as a function of the time after the quench,
for some different values of $g$, starting from the completely ordered
state at $g_0=0$.  To infer the thermodynamic-limit behavior we first
performed numerical simulations at various lattice sizes, up to
$L=24$.  For sufficiently small values of $t$, the curves at the
largest available sizes superimpose, therefore they can be considered
as a good approximation of the thermodynamic large-$L$ limit (see
Fig.~\ref{fig:Ml_t_g1} for an example at $g=1$).  The complete
$L\to\infty$ curve has been then reconstructed by matching with the
analytical asymptotic results discussed above.  Such curves are
reported in Fig.~\ref{fig:Ising_l_gvar_asynt}.  We note that, when
crossing the critical point $g=g_c=1$, they become nonmonotonic and
start oscillating, showing a negative dip.  This may be considered as
a particular signature of the transition point.

We finally mention that Refs.~\cite{RI-11,BRI-12} exploit a
semiclassical framework to compute the longitudinal magnetization and
related correlations.  This semiclassical approach provides reliable
results in regions of the Hamiltonian parameters, where a quasi-particle
picture based on kinks or domain walls turns out to be effective, for
example for small values of the transverse field $g$.

\subsection{Finite-size effects with PBC}
\label{fssdisg00_pbc}

To analyze finite-size effects, we consider again the subtracted
quantity $\delta S$ defined in Eq.~(\ref{submat}), and the
corresponding $\delta M$ for the longitudinal magnetization
\begin{equation}
  \delta M(t,g,L) \equiv M(t,g,L) - {\cal M}(t,g)\,.
  \label{subml}
\end{equation}
Since, for the latter, we do not have the exact large-$L$ curve but
only its large-time exponential behavior ${\cal M}_a$, to evaluate
${\cal M}(t,g)$ we use both numerical results for the largest lattice
sizes supplemented by their asymptotic large-time behavior, as
mentioned at the end of Sec.~\ref{qudisthlim}.

\begin{figure}%[!t]
\includegraphics[width=0.95\columnwidth]{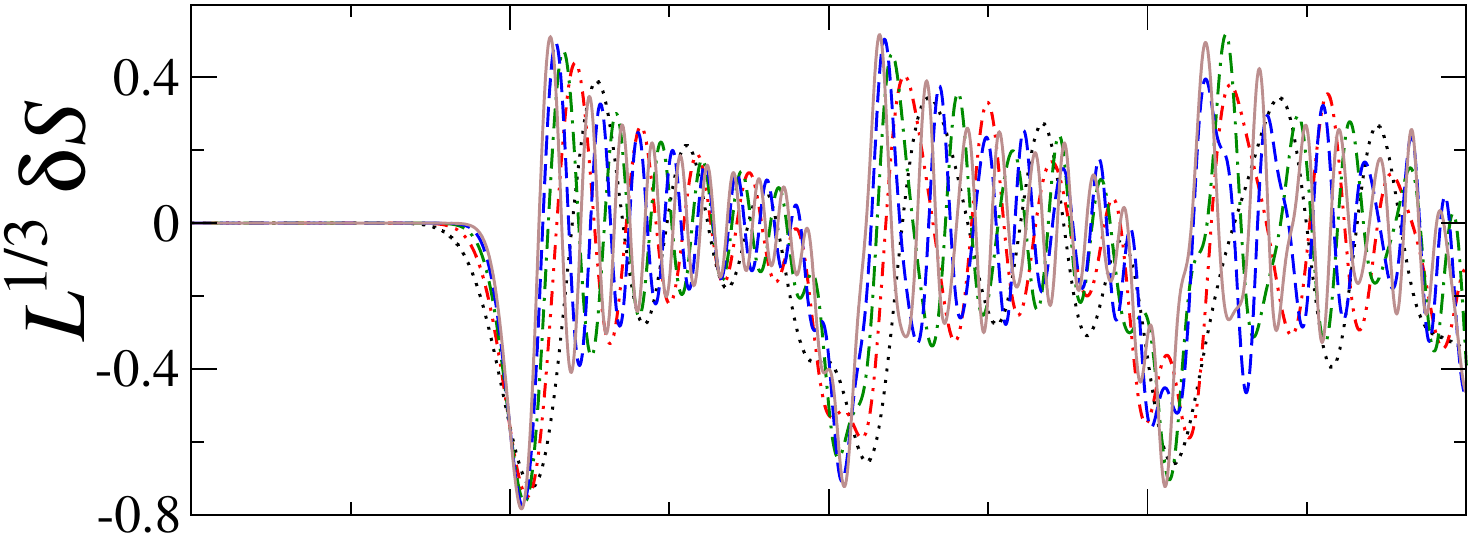}
\includegraphics[width=0.95\columnwidth]{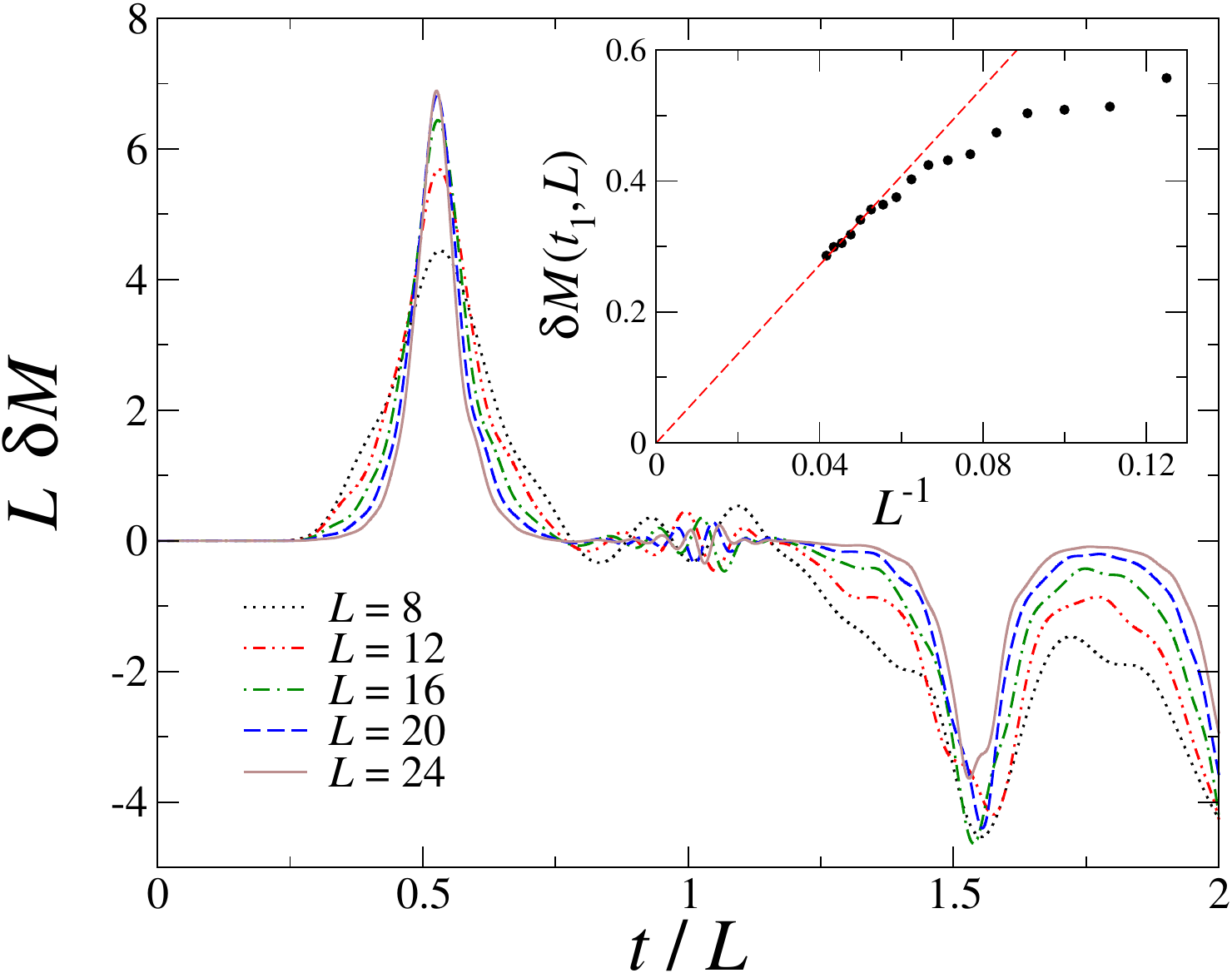}
\caption{Time behavior of the transverse (upper panel) and
  longitudinal (lower panel) magnetization after a quench from the
  symmetry-broken state at $g_0=0$, to $g=1$.  Times have been
  rescaled with $t/L$ and subtracted magnetizations as $L^{1/3} \,
  \delta S$ and $L \: \delta M$, respectively [cf. Eqs.~\eqref{submat}
    and~\eqref{subml}].  The different data sets are for systems with
  PBC and various chain lengths, as indicated in the legend.  The
  inset in the lower panel shows data for $\delta M(t_1, L)$
  vs.~$L^{-1}$; the dashed red line is a linear fit for $L \geq 20$,
  assuming the prediction of Eq.~\eqref{dlmapp} with $\omega = 1$.}
\label{fig:Rescal_ml_g1}
\end{figure}

To begin with, we present results for quenches to the critical point
$g=g_c=1$.  Even in the case of quenches starting from ordered states,
the transverse magnetization behaves consistently with
Eq.~\eqref{asyd}, which explains the finite-size scaling behavior of
the main features of the time dependence, such as the relevant time
scaling variable $t_L\equiv t/L$, the revival features starting at
$t_k = k L/(2 v_m)$, and the $L^{-1/3}$ power law of the envelope of
the short time oscillations. This is witnessed by the curves plotted
in the upper panel of Fig.~\ref{fig:Rescal_ml_g1}, which analyze the
behavior of the transverse magnetization $\delta S$ after a quench
from the perfectly ordered state at $g_0=0$, to the critical point
$g=g_c=1$. Note the appearance of a good data collapse already at $L
\lesssim 24$, after the proper rescaling as predicted in~\eqref{asyd}
using $a= 1/3$ and $t_L = t/L$.

In the lower panel of the same figure we report data of the subtracted
longitudinal magnetization $\delta M$ [cf.~Eq.~\eqref{subml}], for the
same type of quench from the symmetry-broken state at $g_0=0$.  The
subtraction of the asymptotic large-time behavior in the thermodynamic
limit is effective, indeed we clearly observe a relatively large
interval ($t/L \approx 0.2$, in the figure) where the time dependence
appears flat and vanishing.  As expected, the positions of the peaks
and dips confirm that the relevant time scaling variable is $t_L =
t/L$, analogously as for the transverse magnetization (not shown).
However, the behavior of the maxima/minima at the peaks/dips suggests
a different rescaling, as can be seen from the data for the
magnetization values at the first peak vs.~the inverse size $L^{-1}$,
shown in the inset.  Therefore, despite the impossibility to reach
higher sizes with great accuracy prevents us from giving a conclusive
statement, our data hint at the following finite-size scaling behavior
\begin{equation}
  \delta M(t,g=1,L) \approx  L^{-c} f_s(t_L) \,,\qquad c \approx 1\,,
  \label{dlmapp}
\end{equation}
in accordance with the nice approach to data collapse, for the curves
with $L \lesssim 24$ shown in the main frame of the lower panel,
reporting the rescaled magnetization $L\, \delta M$ vs.~the rescaled
time $t/L$.

\begin{figure}%[!t]
\includegraphics[width=0.95\columnwidth]{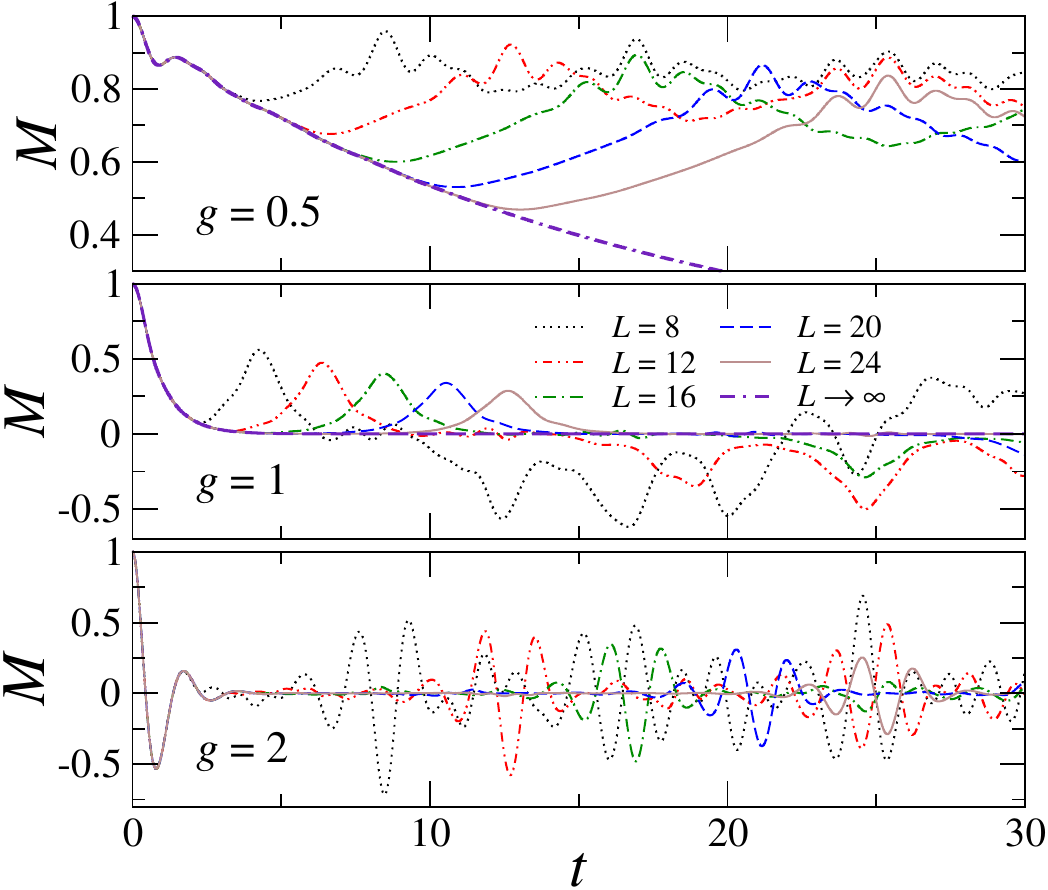}
\caption{Longitudinal magnetization as a function of time, after
  quenches to various values of $g$ in the ordered phase ($g=0.5$,
  upper panel), at the critical point ($g=1$, central panel), and in
  the disordered phase ($g=2$, bottom panel), starting from the fully
  ordered state. The various curves are for different chain lengths,
  as reported in the legend.}
\label{fig:Rescal_ml_gFM}
\end{figure}

The above asymptotic large-$L$ features are also observed in quenches
to values of $g$ different from $g_c=1$, in particular around it. On
the other hand, as shown in Fig.~\ref{fig:Rescal_ml_gFM}, the
situation is less clear for small values of $g$; see e.g.~the case of
$g=0.5$, where the data for different values of $L$ appear to tend to
the same value $M\approx 0.8$.  Of course, this may reflect a slow
convergence in the chain size; in other words, the approach to an
asymptotic behavior such as that reported in Eq.~(\ref{dlmapp}) may be
significantly delayed for quenches to small values of $g$.  Large
lattice sizes would be necessary to clarify this point.

\subsection{Finite-size effects with OBC}
\label{obcsys}

\begin{figure}%[!t]
  \includegraphics[width=0.95\columnwidth]{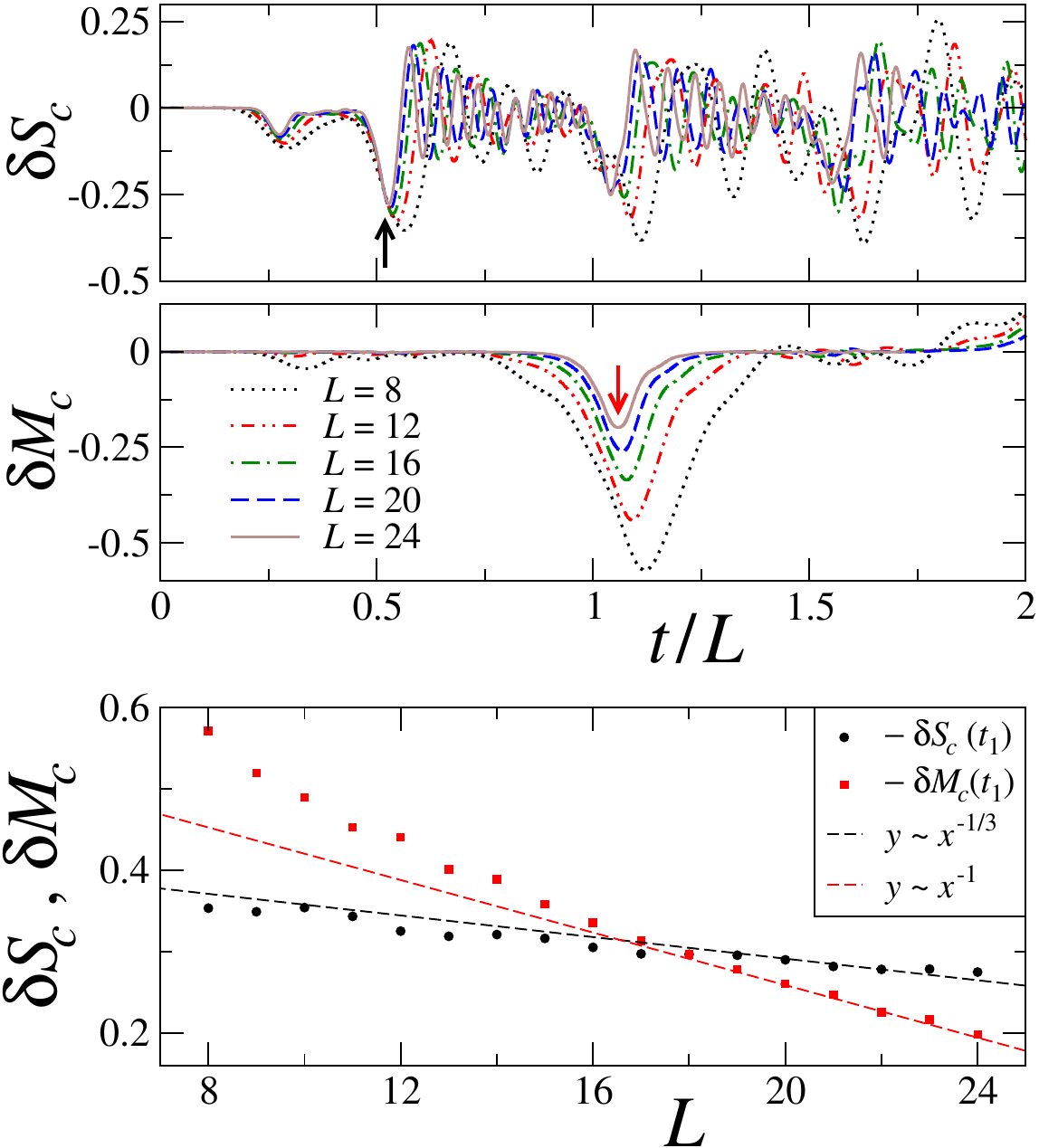}
  \caption{Subtracted transverse (upper panel) and longitudinal
    (central panel) magnetization after a quench starting from the
    symmetry-broken state at $g_0 = 0$, to $g=1$, calculated at the
    center of an Ising chain with OBC. Data are plotted a function of
    the rescaled time $t/L$. The lower panel shows $\delta S_c(t_1)$
    and $\delta M_c(t_1)$ vs.~$L$; the dashed lines indicate the
    behaviors $\delta S_c(t_1) \sim L^{-1/3}$ (black set) and $\delta
    M_c(t_1) \sim L^{-1}$ (red set), respectively.}
\label{fig:Ml_t_g1_OBC}
\end{figure}

Let us finally consider the role of OBC for quenches starting
from the symmetry-broken phase.
The decay of the longitudinal magnetization as a function of the
position, and in particular at the boundaries, was discussed in
Ref.~\cite{IR-11}. Here we only explicitly discuss central
magnetizations, for time evolutions controlled by the critical
Hamiltonian ($g=1$) and starting from the fully ordered state
($g_0=0$).  Data for the subtracted transverse and longitudinal
magnetization, respectively given by Eqs.~\eqref{deltasc} and by the
analogous of Eq.~\eqref{subml} for $M_c$,
\begin{equation}
  \delta M_c(t,g,L) \equiv M_c(t,g,L) - {\cal M}_c(t,g)\,,
  \label{subml_ctr}
\end{equation}
are reported in Fig.~\ref{fig:Ml_t_g1_OBC}.  Of course, the large-$L$
thermodynamic limit is expected to be the same as for $S$ and $M$.
This is confirmed by the results shown in the two upper panels of the
figure, where a plateau equal to zero is clearly visible for short
times $t/L \lesssim 0.2$.
The scaling of first large dip with the size is compatible with what
we already observed for PBC (i.e.~same scaling with $L$, as witnessed
by the data in the bottom panel), cf. Eq.~(\ref{dlmapp}).

\section{Quenches in the ANNNI model}
\label{ANNNI}

We now extend our study to Ising-like systems in the presence of
integrability-breaking perturbations. For this purpose, we consider
quantum quenches in the anisotropic next-nearest neighbor Ising
(ANNNI) chain, changing the quench parameter across an underlying
quantum phase transition. Its Hamiltonian reads
\begin{equation}
  \hat H_{\rm ANNNI} = - J \, 
\sum_{x=1}^L \left[ \hat \sigma^{(1)}_{x\phantom{1}} \hat \sigma^{(1)}_{x+1}
- \kappa\, \hat \sigma^{(1)}_{x\phantom{1}} \hat \sigma^{(1)}_{x+2}
    + g \, \hat \sigma^{(3)}_x \right] \,.
  \label{hannni}
\end{equation}
For values of $\kappa$ sufficiently small, quantum ANNNI models
present a continuous Ising-like transition, see e.g.
Refs.~\cite{BCF-06,HMHPRD-20} and references therein.

\begin{figure}%[!t]
  \includegraphics[width=0.95\columnwidth]{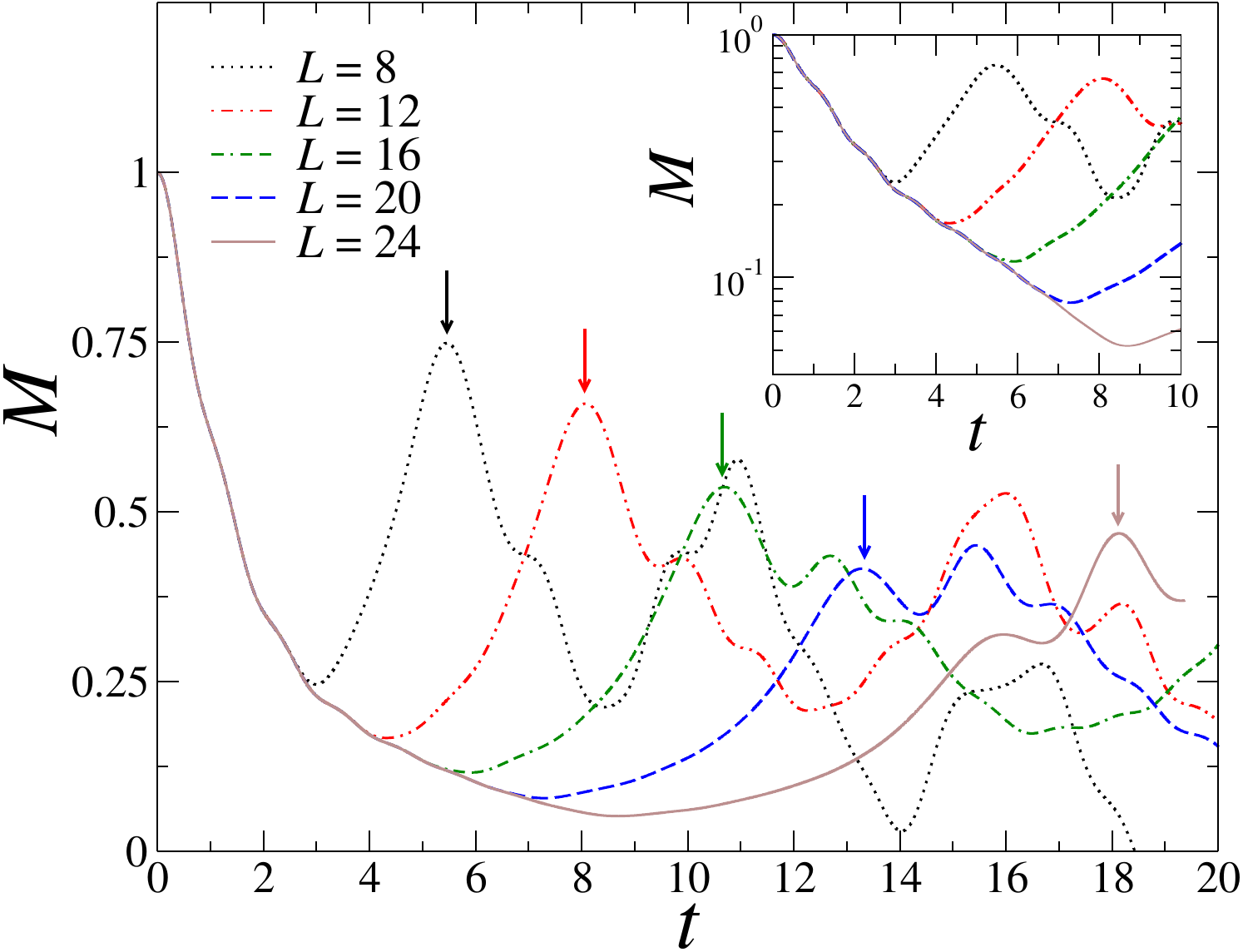}
  \caption{Same as in Fig.~\ref{fig:Ml_t_g1}, but for the post-quench
    critical dynamics of the nonintegrable ANNNI chain with PBC and
    $\kappa = 0.15$.
    Here we set $g=g_c(\kappa=0.15) = 0.73405(4)$~\cite{BCF-06}.}
\label{fig:ANNNI_gc}
\end{figure}

\begin{figure}%[!t]
  \includegraphics[width=0.95\columnwidth]{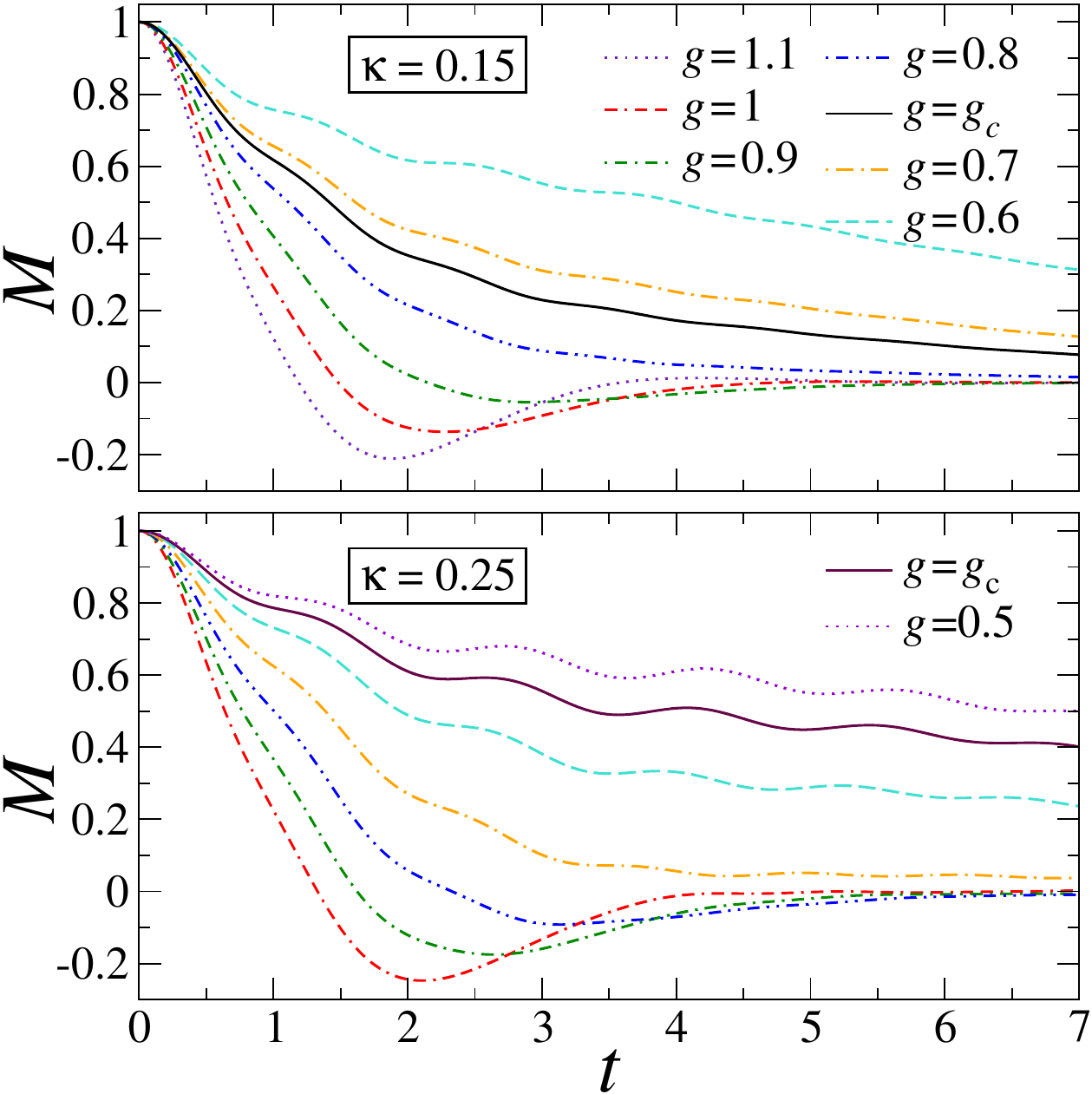}
  \caption{Same as in Fig.~\ref{fig:Ising_l_gvar_asynt},
    but for the ANNNI chain with PBC and $\kappa = 0.15$ (top)
    and $\kappa=0.25$ (bottom). We recall that $g_c = 0.73405(4)$
    for $\kappa=0.15$ and $g_c=0.5403(3)$ for $\kappa=0.25$.}
\label{fig:ANNNI_Ml_Ord}
\end{figure}

In the following we analyze quench protocols starting from the fully
ordered state (corresponding to $g\to 0$) to the critical parameter
$g=g_c(\kappa)$~\cite{CPBF-02,BCF-06}.  Figure~\ref{fig:ANNNI_gc}
displays some results for $\kappa=0.15$, for which~\cite{BCF-06}
$g_c=0.73405(4)$.  They look qualitatively similar to those obtained
for the Ising chain ($\kappa=0$), see Fig.~\ref{fig:Kitaev_t_g1}.  The
results up to $L=24$ allow us to determine the quantum evolution
${\cal M}(t,\kappa,g)$ in the thermodynamic limit up to $t\approx 7$,
by checking their convergence with increasing $L$, see
Fig.~\ref{fig:ANNNI_Ml_Ord}.  The resulting curves in the
thermodynamic limit are shown in Figs.~\ref{fig:ANNNI_Ml_Ord}.

We note remarkable similarities with the case of the Ising chain
($\kappa=0$), in particular the qualitative different behavior of the
evolution of the longitudinal magnetization when quenching to the
ordered and disordered phases.  Indeed quenches within the ordered
phase, $g<g_c$, are characterized by exponential decays to zeroes,
while quenches to the disordered phase, $g>g_c$, show minima, so that
the asymptotic vanishing value gets approached from below.  The
behavior around $g=g_c$ turns out to be less clear, requiring further
and more accurate numerical studies.  This shows that these
qualitative features may persist even in nonintegrable models.  They
provide different characterizations of the different phases, which may
turn useful to distinguish different phases within hard quenching
protocols.  Further studies are deemed in order to understand whether
such qualitative differences can be also used to obtain accurate
estimates of the transition point.

Our analysis substantially supports the idea that the quench dynamics
may be employed to signal the presence of phase transitions even in
nonintegrable systems, as also put forward in Ref.~\cite{HMHPRD-20},
where the peculiar behaviors of some local observables at intermediate
times after quenches were investigated for the purpose of locating the
phase transition.

\begin{figure}%[!t]
  \includegraphics[width=0.95\columnwidth]{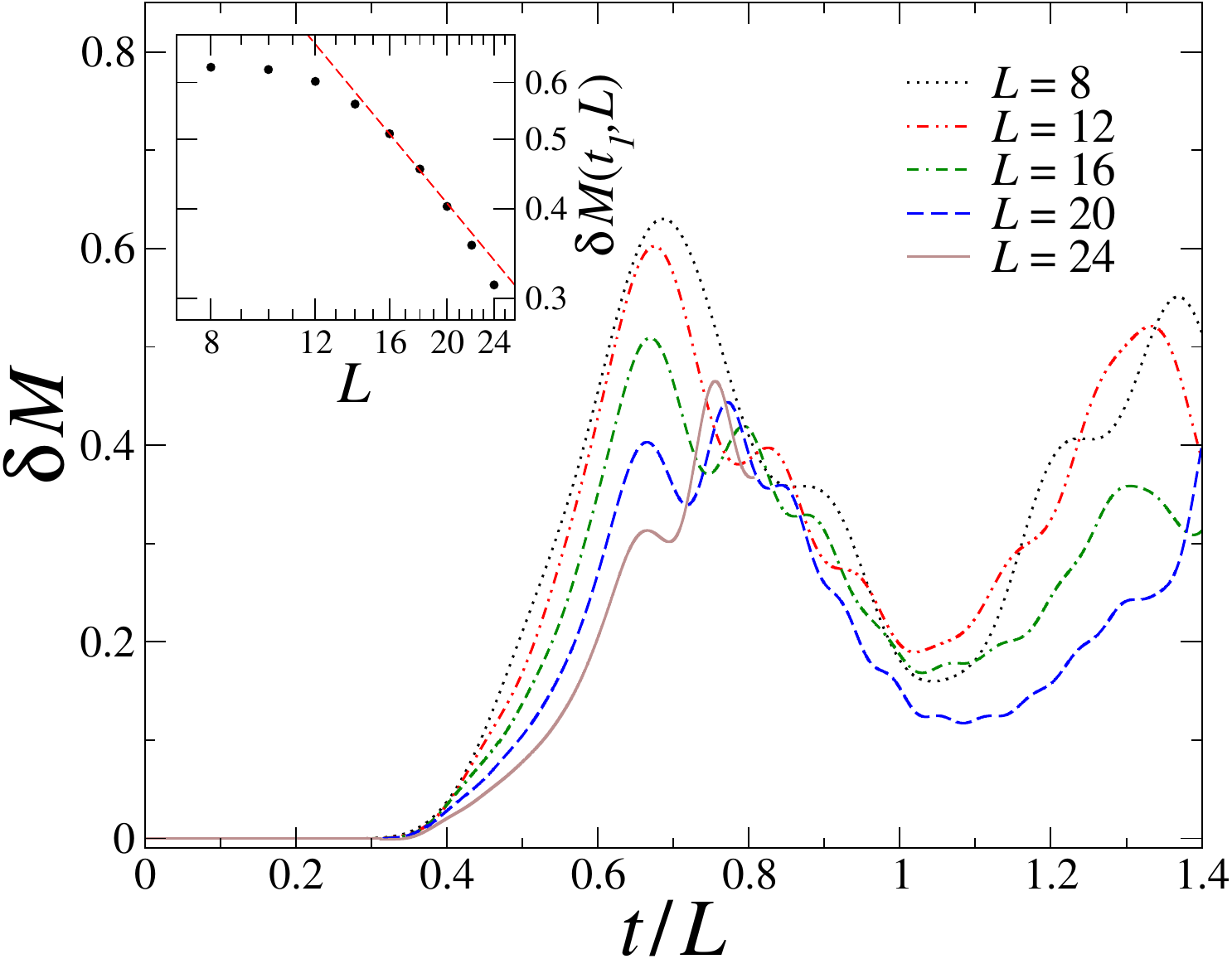}
  \caption{Subtracted longitudinal magnetization after a quench in the
    ANNNI chain with PBC and $\kappa = 0.15$, from the symmetry-broken
    state at $g_0 = 0$, to $g=g_c$.  Data for different system sizes
    are plotted a function of the rescaled time $t/L$. The inset shows
    the value of the subtracted magnetization at the first peak,
    $\delta M(t_1)$, vs.~$L$; the dashed line indicates the behavior
    $\delta M(t_1) \sim L^{-1}$.}
\label{fig:ANNNI_fss}
\end{figure}

Finally, Fig.~\ref{fig:ANNNI_fss} displays the subtracted quantity
\begin{equation}
  \delta M(t,\kappa,g,L) \equiv M(t,\kappa,g,L) - {\cal M}(t,\kappa,g) \,.
  \label{subquaannni}
\end{equation}
for $\kappa=0.15$ and for a quench to $g=g_c$.
Again the behavior is quite similar to that observed for the
Ising chain; in particular, the position of the first peak scales as
$t_L\equiv t/L$. As highlighted in the inset, while the reported data
cannot be considered conclusive due to the small simulatable system
sizes, the decay of $\delta M(t_1)$ is consistent with an asymptotic
$L^{-1}$ behavior, as for the scaling ansatz reported in
Eq.~\eqref{dlmapp}.

\section{Conclusions}
\label{conclusions}

We have investigated the out-of-equilibrium dynamics of
one-dimensional quantum Ising-like systems arising from quenches of
the Hamiltonian parameter $g$ driving the phase transition that
separates the quantum paramagnetic and ferromagnetic phases.  To this
purpose, we have considered the simplest quenching protocol: the
system starts from the ground state of the many-body Hamiltonian
associated with a given value $g_0$ of the parameter; the latter is
then suddenly changed to $g \neq g_0$.  The resulting quantum
evolution is driven either by the unitary dynamics~\eqref{unitary},
for a closed system, or by the Lindblad equation~\eqref{lindblaseq},
in the presence of dissipative interactions with an environment.
The issue we have addressed is whether, and how, quantum
transitions can be probed by the study of such quench protocols.

The behavior of the transverse and longitudinal magnetizations
have been considered both in the infinite-size limit and also
considering finite-size effects, for systems with periodic
and open boundary conditions.
The out-of-equilibrium evolution of such magnetizations in quantum XY
chains, in the thermodynamic limit, cf. Eq.~(\ref{hedefxy}), develops
a singular dependence on the quench parameter $g$ around its critical
value $g_0$, for any anisotropy parameter $\gamma>0$ and starting
point $g_0$, including the extremal ones corresponding to fully
disordered and ordered initial states. Similar singularities have been
reported in Ref.~\cite{BDD-15}.  Finite-size effects,
related to revival phenomena, develop peculiar scaling laws
characterized by power laws. Their exponents
have been accurately determined numerically, suggesting simple
fractions, see e.g. Eqs.~(\ref{asyd}) and~\eqref{dlmapp} for the
transverse and the longitudinal magnetizations, respectively.  Such
power laws are not actually related to the quantum transitions, since
they are observed in generic quenches even those not involving
critical parameters.  They should be related to the interference of
quasi-particle excitations after traveling across the finite chain.
The understanding of such emerging power laws deserve further
investigations.

Finally we have analyzed the effects of two different mechanisms moving
Ising-like systems away from integrability, by adding either
dissipation or further nonintegrable Hamiltonian terms
(such as those of the ANNNI models).  In the case of system-bath
couplings, modeled by a Lindblad equation with local decay and pumping
dissipation operators within the fermionic model obtainable by a
Jordan-Wigner mapping, the singularity of the time evolution of the
transverse magnetization in quenches from disordered states to the
critical Hamiltonian is washed out. Moreover, local dissipation
suppresses finite-size effects exponentially, in particular the
revival phenomena observed in closed systems.  On the other hand, our
analysis of the ANNNI model reveals that some of the main features
of the post-quench dynamics persist, in particular the evolution of the
longitudinal magnetization shows qualitative differences when the
quenches are performed within the ordered phase or crossing the
transition point toward the disordered phase.  Therefore, as already
hinted in Ref.~\cite{HMHPRD-20}, hard quench protocols may be exploited
to get evidence of phase transitions even in nonintegrable systems.

It would be tempting to test whether phenomena similar
to those observed here may emerge also for other classes of quantum
transitions, which include different universality classes
or topological phase transitions.


\begin{thebibliography}{99}

\bibitem{Bloch-08} I. Bloch, Quantum coherence and entanglement with
  ultracold atoms in optical lattices, Nature {\bf 453}, 1016 (2008).

\bibitem{GAN-14}
  I. M. Georgescu, S. Ashhab, and F. Nori, Quantum simulation, Rev. Mod.
  Phys. {\bf 86}, 153 (2014).

\bibitem{Greiner-02} M. Greiner, O. Mandel, T. Esslinger,
  T. W. H\"ansch, and I. Bloch, Quantum phase transition from a
  superfluid to a Mott insulator in a gas of ultracold atoms, Nature
  {\bf 415}, 39 (2002).

\bibitem{Weiss-06} T. Kinoshita, T. Wenger, and D. S. Weiss, A quantum
  Newton's cradle, Nature {\bf 440}, 900 (2006).

\bibitem{Schmiedmayer-07} S. Hofferberth, I. Lesanovsky, B. Fischer,
  T. Schumm, and J. Schmiedmayer, Non-equilibrium coherence dynamics
  in one-dimensional Bose gases, Nature {\bf 449}, 324 (2007).

\bibitem{Trotzky-12} S. Trotzky, Y.-A. Chen, A. Flesch,
  I. P. McCulloch, U. Schollw\"ock, J. Eisert, and I. Bloch, Probing
  the relaxation towards equilibrium in an isolated strongly
  correlated one-dimensional Bose gas, Nat. Phys. {\bf 8}, 325 (2012).

\bibitem{Cheneau-12} M. Cheneau, P. Barmettler, D. Poletti, M. Endres,
  P. Scbau\ss, T. Fukuhara, C. Gross, I. Bloch, C. Kollath, and
  S. Kuhr, Light-cone-like spreading of correlations in a quantum
  many-body system, Nature {\bf 481}, 484 (2012).

\bibitem{Schmiedmayer-12} M. Gring, M. Kuhnert, T. Langen,
  T. Kitagawa, B. Rauer, M. Schreitl, I. Mazets, D. Adu Smith,
  E. Demler, and J. Schmiedmayer, Relaxation and Prethermalization in
  an Isolated Quantum System, Science {\bf 337}, 1318 (2012).

\bibitem{Niemeijer-67}
  Th. Niemeijer,
  Some exact calculations on a chain of spins $1/2$,
  Physica {\bf 36}, 377 (1967);
  Some exact calculations on a chain of spins $1/2$ II,
  Physica {\bf 39}, 313 (1968).

\bibitem{BMD-70} E. Barouch, B. M. McCoy, and M. Dresden, Statistical
  mechanics of the XY model. I, Phys. Rev. A {\bf 2}, 1075 (1970).

\bibitem{DMCF-06}
  G. De Chiara, S. Montangero, P. Calabrese, and R. Fazio,
  Entanglement entropy dynamics of Heisenberg chains,
  J. Stat. Mech. P03001 (2006).

\bibitem{RDYO-07}
  M. Rigol, V. Dunjko, V. Yurovsky, and M. Olshanii,
  Relaxation in a Completely Integrable Many-Body Quantum System:
  An Ab Initio Study of the Dynamics of the Highly Excited States
  of 1D Lattice Hard-Core Bosons,
  Phys. Rev. Lett. {\bf 98}, 050405 (2007).

\bibitem{RDYO-08}
  M. Rigol, V. Dunjko, and M. Olshanii,
  Thermalization and its mechanism for generic isolated quantum systems,
  Nature {\bf 452}, 854 (2008).

\bibitem{ZPP-08}
  M. \v{Z}nidari\v{c}, T. Prosen, and P. Prelov\v{s}ek,
  Many-body localization in the Heisenberg XXZ magnet in a random field,
  Phys. Rev. B {\bf 77}, 064426 (2008).

\bibitem{PZ-09}
  T. Prosen and M. \v{Z}nidari\v{c},
  Matrix product simulations of non-equilibrium steady states
  of quantum spin chains, J. Stat. Mech. (2009) P02035.

\bibitem{PSSV-11}
  A. Polkovnikov, K. Sengupta, A. Silva, and M. Vengalattore,
  Colloquium: Nonequilibrium dynamics of closed interacting quantum systems,
  Rev. Mod. Phys. {\bf 83}, 863 (2011).

\bibitem{IR-11}
  F. Igl\'oi and H. Rieger,
  Quantum Relaxation after a Quench in Systems with Boundaries,
  Phys. Rev. Lett. {\bf 106}, 035701 (2011).

\bibitem{RI-11}
  H. Rieger and F. Igl\'oi,  
  Semiclassical theory for quantum quenches in finite transverse Ising chains,
  Phys. Rev. B {\bf 84}, 165117 (2011).

\bibitem{GS-12}
  A. Gambassi and A. Silva, Large Deviations and Universality
  in Quantum Quenches,  Phys. Rev. Lett. {\bf 109} 250602 (2012).
  
\bibitem{CEF-12-1} P. Calabrese, F. H. L. Essler, and M. Fagotti, Quantum
  quench in the transverse field Ising chain: I. Time evolution of
  order parameter correlators, J. Stat. Mech. (2012) P07016.

\bibitem{CEF-12-2} P. Calabrese, F. H. L. Essler, and M. Fagotti, Quantum
  quenches in the transverse field Ising chain: II. Stationary state
  properties, J. Stat. Mech. (2012) P07022.

\bibitem{BRI-12}
  B. Blass, H. Rieger, and F. Igl\'oi, Quantum relaxation and finite
  size effects in the XY chain in a transverse field after global
  quenches, Eur. Phys. Lett.  {\bf 99}, 30004 (2012).

\bibitem{CE-13}
  J.-S. Caux and F. H. L. Essler,
  Time evolution of local observables after quenching to an integrable model,
  Phys. Rev. Lett. {\bf 110}, 257203 (2013).
  
\bibitem{FCEC-14}
  M. Fagotti, M. Collura, F. H. L. Essler, and P. Calabrese,
  Relaxation after quantum quenches in the spin-1/2 Heisenberg XXZ chain,
  Phys. Rev. B {\bf 89}, 125101 (2014).
  
\bibitem{NH-15}
  R. Nandkishore and D. A. Huse,
  Many body localization and thermalization in quantum statistical mechanics,
  Annu. Rev. Condens. Matter Phys. {\bf 6}, 15 (2015).

\bibitem{CTGM-15}
  A. Chiocchetta, M. Tavora, A. Gambassi, and A. Mitra,
  Short-time universal scaling and light-cone dynamics after a quench
  in an isolated quantum system in $d$ spatial dimensions,
  Phys. Rev. B {\bf 94}, 134311 (2016).

\bibitem{CC-16}
  P. Calabrese and J. Cardy,
  Quantum quenches in $1+1$ dimensional conformal field theories,
  J. Stat. Mech. (2016) 064003.

\bibitem{BD-16}
  D. Bernard and B. Doyon,
  Conformal field theory out of equilibrium: a review,
  J. Stat. Mech. (2016) 064005.

\bibitem{IMPZ-16}
  E. Ilievski, M. Medenjak, T. Prosen, and L. Zadnik,
  Quasilocal charges in integrable lattice systems,
  J. Stat. Mech. (2016) 064008.

\bibitem{LGS-16}
  T. Langen, T. Gasenzer, and J. Schmiedmayer,
  Prethermalization and universal dynamics in near-integrable quantum systems,
  J. Stat. Mech. (2016) 064009.

\bibitem{VM-16}
  R. Vasseur and J. E. Moore,
  Nonequilibrium quantum dynamics and transport: from integrability
  to many-body localization,  J. Stat. Mech. (2016) 064010.
  
\bibitem{NRVH-17}
  A. Nahum, J. Ruhman, S. Vijay, and J. Haah,
  Quantum Entanglement Growth under Random Unitary Dynamics,
  Phys. Rev. X {\bf 7}, 031016 (2017).
  
\bibitem{Heyl-17} M. Heyl, Dynamical quantum phase transitions: a
  review, Rep. Prog. Phys. {\bf 81}, 054001 (2018).

\bibitem{PRV-18b}
  A. Pelissetto, D. Rossini, and E. Vicari, Dynamic finite-size scaling
  after a quench at quantum transitions,  
  Phys. Rev. E {\bf 97}, 052148 (2018).

\bibitem{NRV-19-w} D. Nigro, D. Rossini, and E. Vicari, Scaling properties
  of work fluctuations after quenches near quantum transitions,
  J. Stat. Mech. (2019) 023104.

\bibitem{NRV-19}
  D. Nigro, D. Rossini, and E. Vicari, Competing coherent and dissipative
  dynamics close to quantum criticality, Phys. Rev. A {\bf 100}, 052108 (2019);
  D. Rossini and E. Vicari, Scaling behavior of stationary states arising
  from dissipation at continuous  quantum transitions,
  Phys. Rev. B {\bf 100}, 174303 (2019).

\bibitem{STT-19} J. Surace, L. Tagliacozzo, and E. Tonni, Operator
  content of entanglement spectra in the transverse field Ising chain
  after global quenches, Phys. Rev. B {\bf 101}, 241107(R) (2020).

\bibitem{RV-20-qm}
  D. Rossini and E. Vicari,
  Measurement-induced dynamics of many-body systems at quantum criticality,
  Phys. Rev. B {\bf 102}, 035119 (2020).
  
\bibitem{SGCS-97} S. L. Sondhi, S. M. Girvin, J. P. Carini, and
  D. Shahar, Continuous quantum phase transitions,
  Rev. Mod. Phys. {\bf 69}, 315 (1997).

\bibitem{Sachdev-book}
  S. Sachdev, {\em Quantum Phase Transitions},
  (Cambridge University, Cambridge, England, 1999).

\bibitem{Kibble-76}
  T. W. B. Kibble, 
  Topology of cosmic domains and strings, J. Phys. A {\bf 9}, 1387 (1976).

\bibitem{Zurek-85}
  W. H. Zurek, 
  Cosmological Experiments in Superfluid Helium?,
  Nature {\bf 317}, 505 (1985).

\bibitem{ZDZ-05}
  W. H. Zurek, U. Dorner, and P. Zoller, Dynamics of a quantum phase
  transition, Phys. Rev. Lett. {\bf 95}, 105701 (2005).

\bibitem{PG-08} A. Polkovnikov and V. Gritsev, Breakdown of the
  adiabatic limit in low-dimensional gapless systems,
  Nature Phys. {\bf 4}, 477 (2008).

\bibitem{CEGS-12}
  A. Chandran, A. Erez, S. S. Gubser, and S. L. Sondhi,
  Kibble-Zurek problem: Universality and the scaling limit,
  Phys. Rev. B {\bf 86}, 064304 (2012).

\bibitem{RV-20}
  D. Rossini and E. Vicari, Dynamic Kibble-Zurek scaling framework for 
  open dissipative many-body systems crossing quantum transitions,
  Phys. Rev. Research {\bf 2}, 023211 (2020).

\bibitem{BDD-15} S. Bhattacharyya, S. Dasgupta, and A. Das, Signature
  of a continuous quantum phase transition in non- equilibrium energy
  absorption: Footprints of criticality on higher excited states,
  Sci. Rep. {\bf 5}, 16490 (2015).

\bibitem{RMD-17} S. Roy, R. Moessner, and A. Das, Locating topological
  phase transitions using nonequilibrium signatures in local bulk
  observables, Phys. Rev. B {\bf 95}, 041105(R) (2017).

\bibitem{TIGGG-19}
  P. Titum, J. T. Iosue, J. R. Garrison, A. V. Gorshkov, and Z.-X. Gong, 
  Probing Ground-State Phase Transitions through Quench Dynamics,
  Phys. Rev. Lett. {\bf 123}, 115701 (2019).

\bibitem{HPD-18} M. Heyl, F. Pollmann, and B. Dora, Detecting
  Equilibrium and Dynamical Quantum Phase Transitions in Ising Chains
  via Out-of-Time-Ordered Correlators, Phys. Rev. Lett. {\bf 121},
  016801 (2018).

\bibitem{HMHPRD-20} A. Haldar, K. Mallayya, M. Heyl, F. Pollmann,
  M. Rigol, and A. Das, Signatures of quantum phase transitions after
  quenches in quantum chaotic one-dimensional systems,
  arXiv:2004.02905.

\bibitem{PRV-20} A. Pelissetto, D. Rossini, and E. Vicari, Scaling
  properties of the dynamics at first-order quantum transitions when
  boundary conditions favor one of the two phases,
  Phys. Rev. E {\bf 102}, 012143 (2020).
  
\bibitem{HHH-12} J. H\"app\"ol\"a, G. B. Hal\'asz, and A. Hamma,
  Universality and robustness of revivals in the transverse field XY
  model, Phys. Rev. A {\bf 85}, 032114 (2012).

\bibitem{KLM-14}
  P. L. Krapivsky, J. M. Luck, and K. Mallick,
  Survival of Classical and Quantum Particles in the Presence of Traps,
  J. Stat. Phys. {\bf 154}, 1430 (2014).

\bibitem{Cardy-14}
  J. Cardy, Thermalization and Revivals after a Quantum Quench in
  Conformal Field Theory, Phys. Rev. Lett. {\bf 112}, 220401 (2014).

\bibitem{JH-17}
  R. Jafari and H. Johannesson,
  Loschmidt Echo Revivals: Critical and Noncritical,
  Phys. Rev. Lett. {\bf 118}, 015701 (2017).

\bibitem{MAC-20} R. Modak, V. Alba, and P. Calabrese, Entanglement
  revivals as a probe of scrambling in finite quantum systems,
  J. Stat. Mech. (2020) 083110.

\bibitem{BP-book} H.-P. Breuer and F. Petruccione, {\em The Theory of
  Open Quantum Systems} (Oxford University Press, New York, 2002).
  
\bibitem{CPV-14} M. Campostrini, A. Pelissetto, and E. Vicari,
  Finite-size scaling at quantum transitions, Phys. Rev. B {\bf 89},
  094516 (2014).

\bibitem{Kitaev-01} A. Yu. Kitaev, Unpaired Majorana fermions in
  quantum wires, Phys. Usp. {\bf 44}, 131 (2001).

\bibitem{LR-72}
  E. H. Lieb and D. W. Robinson, The finite group velocity of quantum
  spin systems, Commun. Math. Phys. {\bf 28}, 251 (1972).

\bibitem{CC-05}
  P. Calabrese and J. Cardy, Evolution of Entanglement Entropy in
  One-Dimensional Systems, J. Stat. Mech. (2005) P04010.

\bibitem{Lindblad-76}
  G. Lindblad, On the generators of quantum dynamical semigroups,
  Commun. Math. Phys. {\bf 48}, 119 (1976).

\bibitem{GKS-76}
  V. Gorini, A. Kossakowski, and E. C. G. Sudarshan,
  Completely positive dynamical semigroups of N-level systems,
  J. Math. Phys. {\bf 17}, 821 (1976).

\bibitem{HC-13} B. Horstmann, J. I. Cirac, and G. Giedke, Noise-driven
  dynamics and phase transitions in fermionic systems,
  Phys. Rev. A {\bf 87}, 012108 (2013).

\bibitem{Davies-70} E. B. Davies, Quantum stochastic processes II,
  Commun. Math. Phys. {\bf 19}, 83 (1970); Quantum stochastic
  processes, Commun. Math. Phys. {\bf 15}, 277 (1969).

\bibitem{Evans-77} D. E. Evans, Irreducible Quantum Dynamical
  Semigroups, Commun. math. Phys. {\bf 54}, 293 (1977).

\bibitem{SW-10} S. G. Schirmer and X. Wang, Stabilizing open quantum
  systems by Markovian reservoir engineering, Phys. Rev. A {\bf 81},
  062306 (2010).

\bibitem{Nigro-19} D. Nigro, On the uniqueness of the steady-state
  solution of the Lindblad-Gorini-Kossakowski-Sudarshan equation,
  J. Stat. Mech. (2019) 043202.

\bibitem{Wu-20}
  N. Wu, Longitudinal magnetization dynamics in the quantum Ising ring:
  A Pfaffian method based on correspondence between momentum space and real space,
  Phys. Rev. E {\bf 101}, 042108 (2020).
  
\bibitem{BCF-06} M. Beccaria, M. Campostrini, and A. Feo,
  Density-matrix renormalization-group study of the disorder line in
  the quantum axial next-nearest-neighbor Ising model, Phys. Rev. B
  {\bf 73}, 052402 (2006).

\bibitem{CPBF-02} P. R. Colares Guimar\~aes, J. A. Plascak,
  F. C. S\'a Barreto, and J. Florencio, Quantum phase transitions in the
  one-dimensional transverse Ising model with second-neighbor
  interactions, Phys. Rev. B {\bf 66}, 064413 (2002).

\end{thebibliography}
\end{document}